\documentclass[a4paper,twoside]{article}
\newcommand{\affil}[1]{$^{\rm #1}$}
\usepackage[authoryear]{natbib}
\bibpunct{(}{)}{;}{a}{}{,}
\usepackage{graphicx}
\date{} %Please leave the date blank
\renewcommand{\arraystretch}{1.2}
\usepackage{url}
\usepackage{subfigure}
\usepackage{deluxetable}
\usepackage{pdflscape}

\title{\large\bf\flushleft Characterisation of the Optical Turbulence at Siding Spring}
\author{\parbox{\textwidth}{\flushleft
\vspace{-0.5cm}
{\it Michael Goodwin\affil{A,C,D}, Charles Jenkins\affil{A,E} and Andrew Lambert\affil{B}  } \\
\vspace{0.4cm}
{\small \affil{A}\,Research School of Astronomy  Astrophysics, Australian National University, Mt Stromlo Observatory, via Cotter Rd, Weston, ACT 2611, Australia}\\
{\small \affil{B}\,School of Engineering and Information Technology, UNSW@ADFA}\\
{\small \affil{C}\,Corresponding author. Email: mgoodwin@aao.gov.au}\\
{\small \affil{D}\,Current address: Australian Astronomical Observatory, PO Box 296, Epping, NSW 1710, Australia }\\
{\small \affil{E}\,Current address: Earth Science and Resource Engineering, CSIRO}}}

\begin{document}
\twocolumn[
\begin{changemargin}{.8cm}{.5cm}
\begin{minipage}{.9\textwidth}
\vspace{-1cm}
\maketitle
%
%
%%%%%%%%%%%%%     ABSTRACT    %%%%%%%%%%%%%
%Abstract of no more than 200 words here.
\small{\bf Abstract:}

Measurements of the optical turbulence profile above Siding Spring Observatory were conducted during 2005 and 2006. This effort was largely motivated by the need to predict the statistical performance of adaptive optics at Siding Spring. The data were collected using a purpose-built instrument based on the slope-detection and ranging method (SLODAR) where observations of a bright double star are imaged by Shack-Hartmann taken with the Australian National University 24 inch and 40 inch telescopes. The analysis of the data yielded a model consisting of a handful of statistically prominent thin layers that are statistically separated into the ground layer (37.5, 250~m) and the free atmosphere (1, 3, 6, 9, 13.5~km) for good (25\%), typical (50\%) and bad (25\%) observing conditions. We found that ground-layer turbulence dominates the turbulence profile with up to 80\% of the integrated turbulence below 500~m. The turbulence tends to be non-Kolmogorov, especially for the ground-layer with a power law index of $\beta \sim 10/3$. The mirror/dome seeing can be a significant fraction of the ground-layer turbulence. The median atmospheric seeing, is around 1.2", in agreement with observational reports.

%%%%%%%%%%%%%     KEYWORDS    %%%%%%%%%%%%%
\medskip{\bf Keywords:} site testing --- atmospheric effects --- instrumentation: miscellaneous --- instrumentation: adaptive optics
% Please write all keywords in lower case. PASA uses the
% standard list of subject headings adopted by The Astrophysical Journal
% and available from http://www.journals.uchicago.edu/ApJ/keywords_text.html.
% Keywords are separated by em-dashes, i.e. ---

%%%%%%%%DO NOT EDIT%%%%%%%%%%%%
\medskip
\medskip
\end{minipage}
\end{changemargin}
]
\small
%%%%%%%%EDIT FROM HERE%%%%%%%%%%%%

\section{Introduction}
%Please see the PASA Style Guide for help with correct layout for your manuscript.
%Examples of tables and figures are given below.
%{\em could shorten here, most of the facts are known to your readers}
%Siding Spring Observatory (SSO) is Australia's national astronomical observatory, owned
%and operated by the Australian National University (ANU). The Research School of Astronomy
%\& Astrophysics (RSAA) of the ANU operates the 2.3~m, 40 inch (now closed),
%24 inch (now closed) and new 1.35 m SkyMapper telescopes. The largest telescope at SSO
%is the 3.9 m Anglo-Australian Telescope (AAT) operated by the Anglo-Australian Observatory
%(AAO). The altitude is about 1100 m and SSO is located in the Warrumbungle
%Mountains about 25~km from the small town of Coonabarabran.
%The median astronomical seeing (effective angular resolution) at SSO is about
%1.3 arcsecs, which is more than double the seeing at the best astronomical sites, around
%0.65 arcsecs.

Of interest is the performance of Adaptive Optics (AO) at Siding Spring Observatory (SSO), Australia. The  motivation for  turbulence profiling at SSO is to understand the structure of atmospheric turbulence at a moderate quality astronomical site and to determine the performance predictions for AO. It is know that the performance of AO strongly depends on the structure of the atmospheric turbulence~\citep{Hardy1998}.  No previous detailed site-testing of the structure of atmospheric turbulence (strength and speed) has been undertaken at SSO. The improvements in cost, availability and technology make AO a worthy study at such moderate seeing sites like SSO. AO significantly improves the image quality by compensating for the aberrations induced by atmospheric turbulence in real-time.  The performance of an AO system can be predicted with simulation codes using input atmospheric model optical turbulence profiles (model-OTP) that characterise the turbulence above the astronomical site. It could be that the installation of adaptive optics for the 3.9~m AAT may open the door for new science programs and discoveries that would lead to better science.

If the bulk of the turbulence is low at SSO, then Ground-Layer Adaptive Optics (GLAO) correction mode can provide significant gains.  GLAO provides a  large FoV (6~arcminutes), but with a partial AO correction~\citep{Hubin2006}. Certain science cases (including cosmology and extra-galactic observations) require larger FoVs with excellent seeing conditions, which can be achieved with GLAO for a larger fraction of nights~\citep{Hubin2006}. The GLAO correction mode averages the wavefront from several or more widely separated wavefront sources and applies the result to a single deformable mirror conjugated at the ground. Hence GLAO performance is typically best when the bulk of turbulence is near the ground. It is known that the bulk fraction of atmospheric turbulence at most astronomical sites is located near the ground~\citep{Hardy1998}, in the ``boundary layer" below 500~m in altitude. The figure of merit for the GLAO correction mode is the Ensquared Energy (EE) fraction in a pixel. In some cases the improvement in Ensquared Energy can be more than double the natural seeing and hence halves the integration times to achieve the same signal-to-noise ratios~\citep{Hubin2006}.

Hence, the observations of turbulence profiles are of critical importance as the statistical analysis reveals a set of model optical
turbulence profiles that serve as input atmospheres into AO simulation codes. This fact has provided the
necessary justification for turbulence ranging campaigns at major astronomical sites as
well as research into various site-testing instruments.

The characteristic structure of the atmospheric turbulence above SSO was not well understood
prior to our site-testing campaign. However, the seeing resulting
from the total turbulence integral is better understood with DIMM seeing measurements
at SSO reported by~\cite{Wood1995}. The common seeing values reported by \cite{Wood1995} are ∼ 1.2".
Several turbulence profiles have been observed above SSO (January 1997) using the Generalized
SCIDAR method with the ANU 40" telescope and are reported by \cite{Klueckers1998} where is is noted
that the strongest turbulent layers are located near the ground in the so called ‘boundary-layer’
with heights below 3 km. However, a statistically robust model-OTP cannot be derived due
to an insufficient number of profiles observed by \cite{Klueckers1998}.

This paper discusses the $C_N^2(h)$  and $V(h)$ profile measurements taken at SSO and the derived model-OTP
for suitable use in AO simulation codes. Section 2 introduces the turbulence parameters. Section 3 outlines the technique and instrumentation
used in the measurement of the $C_N^2(h)$ and $V(h)$ profiles. Section 4 and 5 discusses the data and trends noted in the measured profiles.
Section 6 introduces the model-OTP that characterises the atmospheric turbulence profile at SSO. Concluding remarks are in section 7.

The  predicted performance of AO at SSO based on the presented model-OTP will be published in a forthcoming paper.

\section{Turbulence Parameters}

Atmospheric turbulence exhibits a physical process that is complex and random in nature,
requiring a suitable model. A widely accepted model is that proposed by Kolmogorov~\citep{Hardy1998},
who investigated the mechanical structure of atmospheric turbulence.
The Kolmogorov model described the velocity of motion in a fluid medium, where
energy is added in the form of large-scale disturbances which then break down to smaller
and smaller structures, until an inner-scale is reached.

From the Kolmogorov model, a spatial power spectrum of phase (power law index, $\beta=11/3$) can be deduced followed by a
set of structure functions, that describe non-stationary random fluctuations encountered
by atmospheric turbulence. Using these structure functions, a set of general turbulence
parameters can be specified that summarize the effects of atmospheric turbulence. The
key parameter used in the calculation of the turbulence parameters is the refractive index
structure constant, $C_N^2$ (units $m^{-2/3}$), and its variation with altitude, $z$, and time, $t$. We
now proceed with the specification of the most useful turbulence parameters for AO.

The full width half maximum (FWHM) seeing angle or image dispersion for long exposure images, $\phi$ (rad), is given by:

\begin{equation}
\phi  = \frac{\lambda }
{{r_0 }}
\end{equation}

where $\lambda$ (m) is the wavelength and $r_0$ (m) is the coherence length ~\citep{Fried1966}. Partial AO compensation results in a diffraction limited core surrounded by a broader halo equal to the seeing disk,  $\phi$. The parameter $r_0$ includes the integrated effect of the refractive index fluctuations over the vertical propagation path, $z$ (m). It represents a fictitious ``cell size" of turbulence and defines an aperture diameter over which the mean-square wavefront error is 1~rad$^2$ given by:

\begin{equation}
\label{eqn:rnaughtintro}
r_0  = 0.185\lambda ^{\tiny{{\raise0.7ex\hbox{$6$} \!\mathord{\left/
 {\vphantom {6 5}}\right.\kern-\nulldelimiterspace}
\!\lower0.7ex\hbox{$5$}}}} \left( {\sec \zeta } \right)^{\tiny{{\raise0.7ex\hbox{${ - 3}$} \!\mathord{\left/
 {\vphantom {{ - 3} 5}}\right.\kern-\nulldelimiterspace}
\!\lower0.7ex\hbox{$5$}}}} \left[ {\int\limits_z {C_n^2 (z)dz} } \right]^{ - \frac{3}
{5}}
\end{equation}

where $\zeta$ (rad) is the zenith angle. The coherence length, $r_0$, also corresponds to the approximate spatial scale that adaptive optics must measure and compensate the effects of atmospheric turbulence. The angular displacement over which the mean-square wavefront error is 1~rad$^2$ is called the isoplanatic angle ~\citep{Fried1982}, $\theta_0$ (rad), given by:

\begin{equation}
\label{eqn:isoangle}
\theta _0  = 0.058\lambda ^{\tiny{{\raise0.7ex\hbox{$6$} \!\mathord{\left/
 {\vphantom {6 5}}\right.\kern-\nulldelimiterspace}
\!\lower0.7ex\hbox{$5$}}}} \left( {\sec \zeta } \right)^{\tiny{{\raise0.7ex\hbox{${ - 8}$} \!\mathord{\left/
 {\vphantom {{ - 8} 5}}\right.\kern-\nulldelimiterspace}
\!\lower0.7ex\hbox{$5$}}}} \left[ {\int\limits_z {C_n^2 (z)z^{\tiny{{\raise0.7ex\hbox{$5$} \!\mathord{\left/
 {\vphantom {5 3}}\right.\kern-\nulldelimiterspace}
\!\lower0.7ex\hbox{$3$}}}} (z)dz} } \right]^{ - \frac{3}
{5}}.
\end{equation}

The isoplanatic angle, $\theta_0$, can be considered the approximate compensated field of view for on-axis single-conjugate adaptive optics; or the maximum allowable angular distance from the high-order wavefront source to the science object. The coherence time ~\citep{Greenwood1977}, $\tau_0$ (s), is roughly the time taken for the wind to move turbulence by $r_0$. This is given by:

\begin{equation}
\tau _0  = 0.058\lambda ^{\tiny{{\raise0.7ex\hbox{$6$} \!\mathord{\left/
 {\vphantom {6 5}}\right.\kern-\nulldelimiterspace}
\!\lower0.7ex\hbox{$5$}}}} \left( {\sec \zeta } \right)^{\tiny{{\raise0.7ex\hbox{${ - 3}$} \!\mathord{\left/
 {\vphantom {{ - 3} 5}}\right.\kern-\nulldelimiterspace}
\!\lower0.7ex\hbox{$5$}}}} \left[ {\int\limits_z {C_n^2 (z)v_{wind} ^{\tiny{{\raise0.7ex\hbox{$5$} \!\mathord{\left/
 {\vphantom {5 3}}\right.\kern-\nulldelimiterspace}
\!\lower0.7ex\hbox{$3$}}}} (z)dz} } \right]^{ - \frac{3}
{5}}.
\end{equation}

The coherence time, $\tau_0$, can be considered as the maximum duration that the atmospheric turbulence can be considered 'frozen'; or the maximum duration allowable between sequential wavefront samples and corrections for the adaptive optics control system.

Common site-testing instruments that measure atmospheric turbulence usually assume a Kolmogorov model of turbulence with power law index, $\beta=11/3$. All of these parameters can also be measured in the case of non-Kolmogorov turbulence but they all become functions of $\beta$. To estimate the performance of optical systems in non-Kolmogorov turbulence, the power spectral density can be expressed~\citep{Stribling1995} as

\begin{equation}
\Phi _n (\kappa ,\beta ,z) = a(\beta )B(z)\kappa ^{ - \beta },
\end{equation}
where $\Phi _n (\kappa ,\beta ,z)$ is the power spectral density as a function of position, $z$ is along the optical path, $\beta$ is the power law slope (11/3 for Kolmogorov), $B(z)$ is the index structure constant having units $m^{3-\beta}$ and  $a(\beta )$ is a function to maintain consistency with the index structure function and is given by ~\cite{Stribling1995} as

\begin{equation}
D_n (r) = B(z)r^{\beta  - 3}.
\end{equation}

The analysis of the non-Kolmogorov model is appropriate given that the value of the power law index, $\beta$, can be determined by the function fitting SLODAR method ~\citep{Butterley2006}.

%%%%%%%%%%%%%%%%%%%%

\section{Turbulence Measurement}

Various methods are used for turbulence profiling, including direct
sensing with microthermal sensors on towers~\cite{Pant1999} or balloons
~\cite{Azouit2005}, remote-sensing with acoustic scattering (SODAR) ~\cite{Travouillon2006}, or triangulation of scintillation (SCIDAR)~\citep{Vernin1973,Fuchs1994} or of image motion (SLODAR) ~\citep{Wilson2002,Butterley2006,Goodwin2007}. These techniques have reached a degree of maturity exhibiting reasonable agreement when used together in campaigns (\cite{TokovininTravouillon2006}, Cerro Tololo campaign \citep{Sarazin2005}). Each technique has its unique benefits and limitations in terms of cost, height resolution, height range, temporal resolution, ease of implementation and data reduction complexity.

\subsection{SLODAR Method}

The SLODAR (Slope Detection And Ranging) technique has been used on large telescopes at the ORM, La Palma and later on  a portable, stand alone, turbulence profiler for ESO, based on a 40~cm telescope with an $8\times8$ Shack-Hartmann Wavefront Sensor (SHWFS)  (5~cm sized sub-apertures)~\citep{Wilson2002}. The larger apertures 5 to 15~cm of SLODAR relaxes exposures times to 4 to 8~ms (typical wind crossing timescales) providing more suitable observational targets. The ground-layer can be measured with sufficient height resolutions (50 to 100~m) by observing widely separated double stars~\citep{Wilson2002}  whereas higher altitudes can be investigated by observing more narrowly separated doubles. For these reasons, we have selected the SLODAR method for our site-testing campaign to measure and characterise the turbulence profiles at SSO.

\begin{figure} [h]
\centering
\includegraphics[width=\columnwidth]{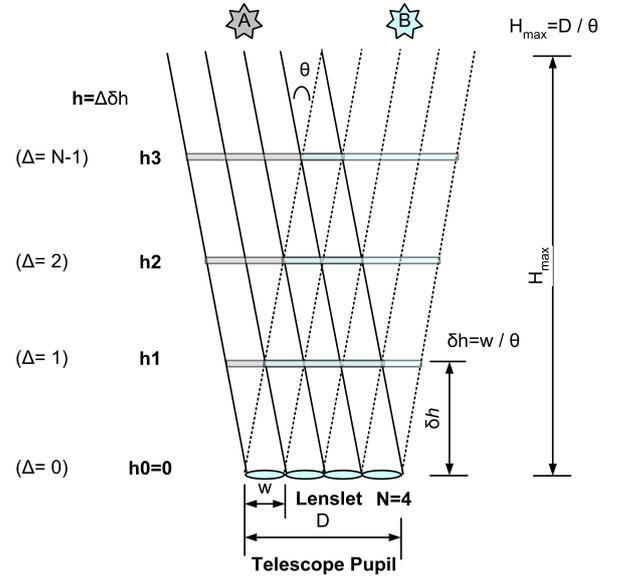}
  \caption[Schematic of the SLODAR geometry.]{Diagram illustrating the geometry of the SLODAR method for a N=4 system. $\theta$ is the double star angular separation. $D$ is the diameter of the telescope pupil and $w$ the width of the sub-aperture of the Shack-Hartmann Wavefront Sensor (SHWFS) array. The centers of the altitude bins are given by $\Delta \delta h$ where $\Delta$ is the lateral pupil separation (units of $w$) and $\delta h = w/\theta$. The ground-layer can be analyzed in higher-resolution by utilization of double stars having larger $\theta$. }\label{fig:SLODAR_Geometry}
\end{figure}

The SLODAR method, illustrated in Figure~\ref{fig:SLODAR_Geometry}, is an optical triangulation method with the turbulence information extracted from the cross-covariance of the wavefront slopes of a double star measured using a Shack-Hartmann Wavefront Sensor (SHWFS). The two optical wavefront components of the double star with separation $\theta$, passing through a single turbulent layer, at altitude $H$, produce copies of the aberrations at the telescope pupil that are displaced by $S=H \theta$ along the axis of the double star separation. The corresponding turbulent layer shows up in the spatial cross-covariance of the optical wavefronts at a spatial offset $S$.

The SHWFS measures the wavefront slopes by optically dividing the telescope pupil into an ($N \times N$) array of square sub-apertures, accompanying each lens in a microlens array and measuring the centroids of the spot displacements, being proportional to the averaged wavefront slope. The sub-apertures and the detector have a sufficient field of view to measure the ($N \times N$) array of spot patterns from both components of the double star simultaneously. The exposure times are typically 4- 8~ms to freeze the turbulence, being directly proportional to sub-aperture size, $w$, related to the wind speed, $v$, with crossing timescales, $\tau=w/v$. The sub-aperture sizes are designed to be approximately equal to or less than $r_0$, or ranging from 5~cm (poor seeing) to 15~cm (good seeing) depending on the median seeing.

The height resolution is uniform, given by $\delta h = w / \theta $ (at zenith). The highest sampled layer, $h_{N-1}=(N-1)\delta h \approx H_{max} = D / \theta$, where $N$ is the number of sub-apertures across the telescope pupil, with the ground layer denoted as $h_{0} = 0$ with resolution $\delta h / 2$. The vertical resolution and maximum sample height are scaled by the inverse of the air mass, $\chi$, or $\cos(\zeta)$, where $\zeta$ is the zenith distance.

\subsection{SLODAR Instrument}

The SLODAR site-testing campaign to characterise the atmospheric turbulence above SSO consists of results that
were obtained from 8 one-week observing runs spanning years 2005 to 2006 with the purpose-built 7$\times$7
(Runs 1-6) and 17$\times$17 (Runs 7-8) SLODAR instrument configurations using the ANU 24" and 40" telescopes.

\begin{figure*}
\centering
  \includegraphics[width=0.8\textwidth,bb=0 0 890 263]{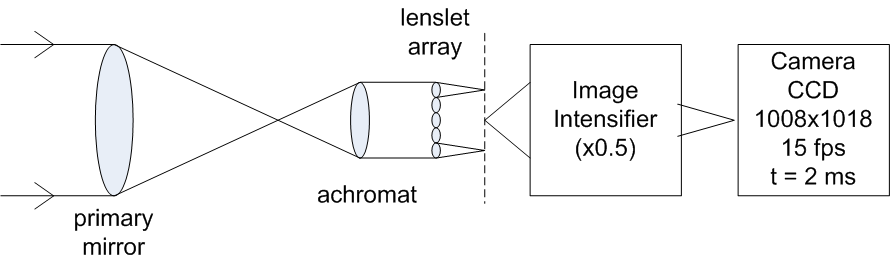}\\ % slodarinstrument1.eps
  \caption[SLODAR instrument schematic with expanded optical diagram]{ The expanded optical block diagram of the  $7\times7$ SLODAR instrument (first version) that attaches via bayonet mount to the focus of the ANU 24 inch telescope at SSO. Diagram shows $5\times5$ SHWFS for simplification. }\label{fig:slodarinstrument1}
\end{figure*}

\begin{figure}
\centering
  \includegraphics[width=\columnwidth]{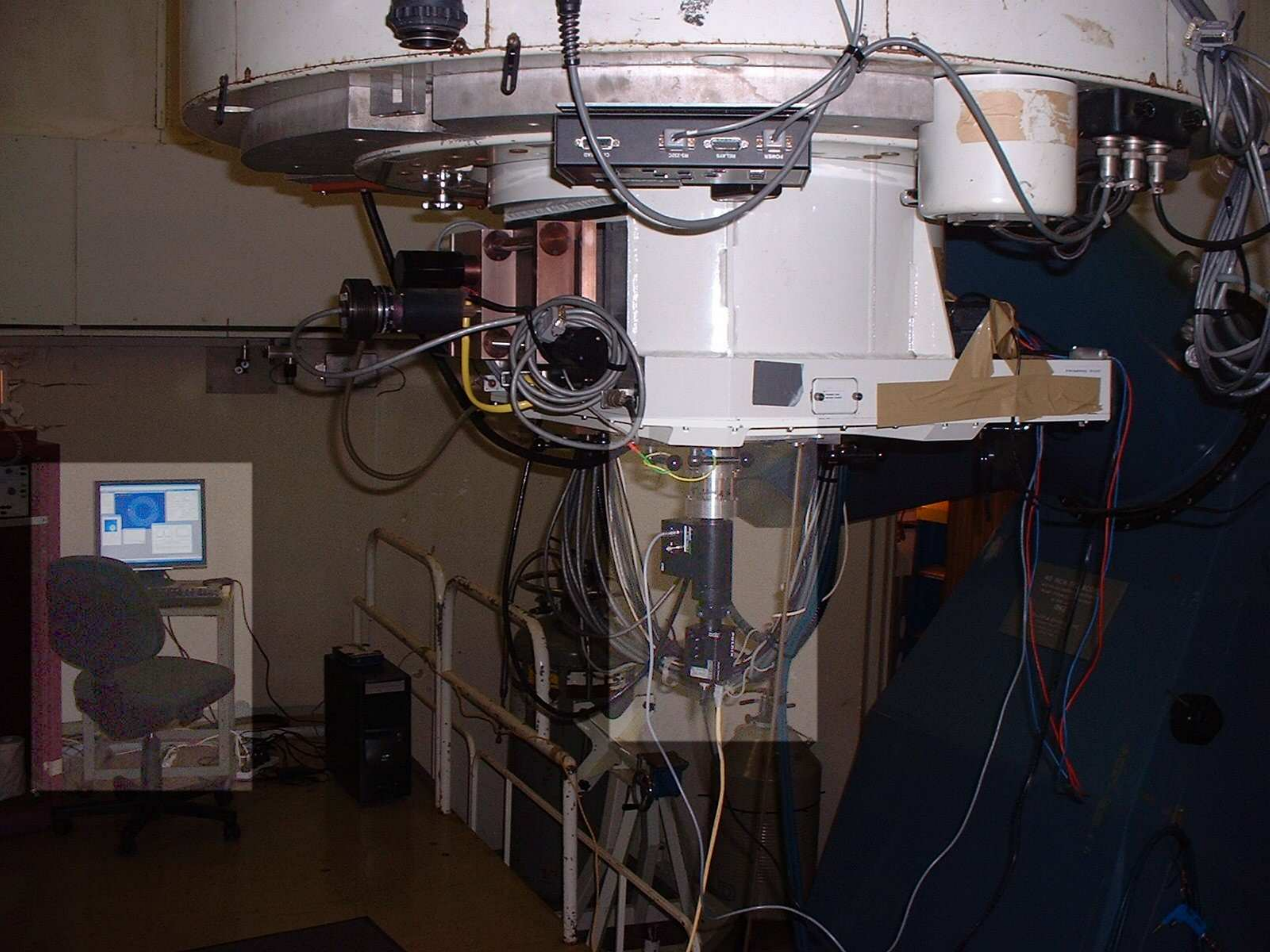}\\
  \caption[ANU 17$\times$17 SLODAR Instrument] {The ANU 17$\times$17 SLODAR Instrument (5.8~cm sub-apertures) on the ANU 40" telescope at SSO (Photograph taken 18 June 2006).} \label{fig:systemtelescope1}
\end{figure}

The SLODAR instrument can be compared to that of a SHWFS as used in adaptive optics to measure the aberrated optical wavefront. However, the SHWFS of the SLODAR instrument has a much wider field of view requiring the simultaneous measurement of double star optical wavefront gradients. The double stars observed by SLODAR can have angular separations up to several arcminutes. A functional diagram of the first ANU $7\times7$ SLODAR instrument installed on the ANU 24" telescope at SSO is shown in Figure~\ref{fig:slodarinstrument1}. The optical diagram for the 17$\times$17 SLODAR instrument on the  ANU 40" telescope is similar to that shown in Figure~\ref{fig:slodarinstrument1} except for the 2$\times$ optic before the image intensifier. The 17$\times$17 SLODAR instrument is shown in Figure~\ref{fig:systemtelescope1}.

An advantage of the SLODAR instrument is a relatively simple optical design. A single collimator lens images the telescope pupil onto the  (Microlens Array) MLA as shown in Figure~\ref{fig:slodarinstrument1}. The collimating lens performs the following functions: (i) de-magnifies and collimates the incoming light beams of each double star component (i.e. point source at infinity); (ii) re-images the telescope pupil onto the MLA.

The MLA performs the function of optically dividing the telescope pupil into an array of sub-apertures. Each sub-aperture forms two spots (image of double star) at their focal plane with image motion caused by aberrations in the double star optical wavefronts. The width of the sub-apertures, $w$, are comparable to the seeing coherent length, $r_0$, keeping the aberrated optical wavefront approximately linear. The sub-aperture spot displacement from mean position (centroid) is proportional to the averaged wavefront gradient (or slope). The images of the telescope pupils formed at the MLA are completely overlapped and are produced by double star components $A$ and $B$. The individual lenslets (or sub-apertures) typically have slow focal ratios that image double star $A$ and $B$ components onto the image intensifier input (photocathode).

The image intensifier performs the function of applying a high gain ($>1000$) to the incoming signal photons overcoming the high read noise of the high-frame rate cameras used with the SLODAR instruments. The image intensifier allows fainter double stars to be observed, down to limiting magnitudes in the V-band of approximately 5 to 6 (compared to 1 to 2 without the image intensifier). The image intensifier re-images and de-magnifies the MLA spot patterns onto the camera detector. Hence the SLODAR instrument has two focal planes requiring correct adjustments during the calibration process.

The camera detectors used with the SLODAR instruments have the key features (i) large format cameras (e.g. 1018$\times$1008 pixel array) to image wide double stars; (ii) high frame rates (15, 20, 30, 200~fps) for layer wind speed measurements; (iii) short camera exposures of 2~ms to 8~ms in order to `freeze' the turbulence. The camera exposures are on time scales equivalent to the turbulent layer wind crossing times of the sub-apertures.

Complementing the 17$\times$17 instrument was the purpose-built real-time software,  (a graphical user interface (GUI) application) to satisfy the requirements for instrument control, processing and diagnostics and autonomously logging of centroid data. The real-time software significantly increased the number of quality observed datasets.

%%%%%%%%%%%%%%%%%%%%

\section{Data Description}

\subsection{Observational Plan}

The scheduled observing runs for the SSO turbulence profiling are listed in Table~\ref{tab:slodarrunssso}. The observational plan was to sample the turbulence profile for each season over the course of the whole year. Turbulence profiling results were obtained from 8 one-week observational runs spanning years 2005 to 2006 with the  $7\times7$ (Runs 1-6) and $17\times17$ (Runs 7-8) SLODAR instruments using the ANU 24" and 40" telescopes respectively. Sampling over three consecutive months was conducted from November 2005 to January 2006 to examine any possible monthly variation in turbulence profiles. The first observational run in May 2005 provided a test run of the  7$\times$7 SLODAR instrument on the ANU 24" telescope. A total of six observing runs were conducted with the ANU 7$\times$7 SLODAR instrument, the final run being in January 2006. The instrument changeover for the observational run in April 2006 to the 17$\times$17 SLODAR instrument on the ANU 40" telescope was a response to the scientific requirement of needing more height sampling resolution bins and the measurement of the turbulent layer wind speeds.

\begin{table*}[tbp]
\begin{center}
\caption{SLODAR observational runs at SSO}\label{tab:slodarrunssso}
\begin{tabular}{lllll}
\hline Run & Scheduled Dates & Nights & Profiles & Telescope/Instrument \\
\hline 1 & 2-8 May 2005 & 4 of 7 & - & ANU 24" /  7$\times$7 SLODAR  \\
 2 & 14-20 June 2005 & 3 of 7 & 31 & ANU 24" /  7$\times$7 SLODAR  \\
 3 & 21-27 September 2005 & 3 of 7 & 86 & ANU 24" /  7$\times$7 SLODAR  \\
 4 & 1-7 November 2005 & 1 of 7 & 16 & ANU 24" /  7$\times$7 SLODAR  \\
 5 & 12-18 December 2005 & 2 of 7 & 37 & ANU 24" /  7$\times$7 SLODAR  \\
 6 & 18-24 January 2006 & 5 of 7 & 136 & ANU 24" /  7$\times$7 SLODAR  \\
 7 & 11-17 April 2006 & 5 of 7 & 450 & ANU 40" /  17$\times$17 SLODAR  \\
 8 & 15-21 June 2006 & 6 of 7 & 1892 & ANU 40" /  17$\times$17 SLODAR  \\
\hline
\end{tabular}
\medskip\\
%$^a$Table footnotes go here.\\
\end{center}
\end{table*}

The observational list of double star targets are tabulated in Table~\ref{tab:slodartargets}. The limiting magnitude of the instruments, V $\sim$ 5.5, limited the observational list to a couple (on occasions only one) of suitable targets for any given time at the telescope. The ability to alternate between ground-layer and free-atmosphere sampling is facillitated by switching between widely to narrowly separated double star targets. The increased number of profiles (datasets) for the final two observing runs was a result of changing from the manual process of logging data to automated logging with the introduction of the real-time software. The final run in June 2006 witnessed the full operation of the real-time software which captured 1892 datasets for off-line processing.

\begin{table*}[tbp]
\begin{center}
\caption{Double star targets for SLODAR observations at SSO. }\label{tab:slodartargets}
\begin{tabular}{lllll}
\hline {Name} & {$\alpha$ (RA)} & {$\delta$ (DEC)} & {mag} &  {separation} \\
\hline $\theta$ Eri & 02 58 & -40 18 & 3.4/4.5 & 8.2 ''  \\
DUN 16 Eri   & 03 49 & -37 37 & 4.8/5.3 & 7.9 ''  \\
$\theta$ Ori & 05 35 & -05 25 & 4.9/5 & 135 ''  \\
$\beta$ Mon & 06 28 & -07 02 & 4.7/5.2 & 7.2 ''  \\
$\alpha$ Cru & 12 26 & -63 06& 1.3/1.8 & 4.3 ''  \\
$\mu$ Cru    & 12 55 & -57 11 & 4.0/3.5 & 34.9 ''  \\
HIP65271     & 13 23 & -60 59 & 4.5/6.1 & 60 ''  \\
$\alpha$ Cen & 14 39 & -60 50 & -0.01/1.33 & 9.5 ''  \\
$\theta$ Ser & 18 56 & +04 12 & 4.6/4.9 & 21.6 ''  \\
$\delta$ Aps & 16 20 & -78 42 & 4.7/5.3 & 102.9 ''  \\
\hline
\end{tabular}
\end{center}
\end{table*}

\subsection{Data Acquisition}

\begin{figure}[htbp]
  \begin{center}
    \mbox{
      \subfigure[]{\scalebox{1.0}{\includegraphics[width=0.9\columnwidth, bb=0 0 512 496] {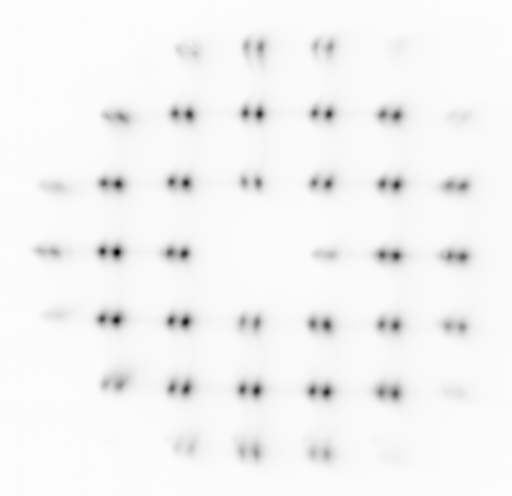}}}
      }
    \mbox{
      \subfigure[]{\scalebox{1.0}{\includegraphics[width=0.9\columnwidth, bb=0 0 419 401] {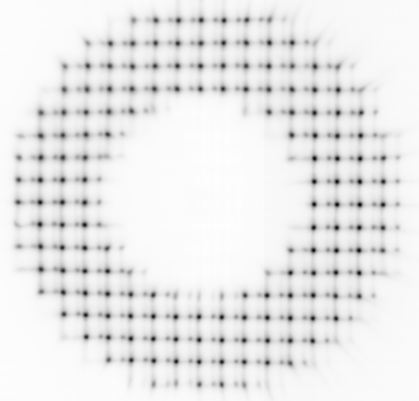}}}
      }
    \caption{An ensemble average over raw camera frames of the (a) SLODAR 7$\times$7 Instrument on ANU 24" telescope at SSO (b) SLODAR 17$\times$17 Instrument on ANU 40" telescope at SSO. The double stars observed are (a) $\alpha~Crux$ and (b) $\alpha~Cen$  are aligned along the SHWFS $x$-direction.} \label{fig:sso_shwfs_data}
  \end{center}
\end{figure}

The SLODAR instrument delivers raw camera frames as shown in Figure~\ref{fig:sso_shwfs_data} and are processed either by MATLAB$\textregistered$ \citep{Matlabref} (ANU 7$\times$7 SLODAR instrument) or by the SLODAR real-time software (ANU 17$\times$17 SLODAR instrument). The SLODAR real-time software has the capability to log raw camera frames but typically only the centroid data is logged.

A summary of the observational data statistics are listed in Table~\ref{tab:slodarrunssso}. Data is acquired at 15~fps (Pulnix$\textregistered$ \citep{PulnixRef} TM1020, 1018$\times$1008 pixels) and 30~fps (Pixelink$\textregistered$ \citep{PixelinkRef} A741, 1280$\times$1024 pixels) to image larger separated double stars and 200~fps (Pulnix TM6740GE, 640$\times$480 pixels) for high temporal sampling. A typical dataset consists of a minimum of 600 camera frames (15~fps) or 4000 camera frames (200~fps) for sufficient statistical sampling of the atmospheric turbulence.

%%%%%%%%%%%%%%%%%%%%

\section{Data Analysis}

\subsection{$C_N^2(h)$  profiles}
\label{sec:turbulencedistribution}
\begin{figure*}[]
  \begin{center}
    \mbox{
      \subfigure[]{\scalebox{1.0}{\includegraphics[width=0.45\textwidth, bb=0 0 560 420]{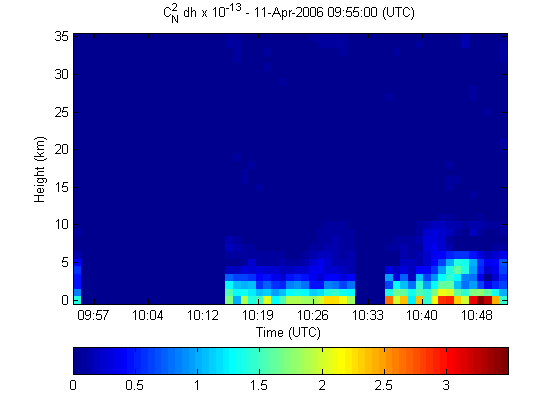}}} \quad
      \subfigure[]{\scalebox{1.0}{\includegraphics[width=0.45\textwidth, bb=0 0 560 420]{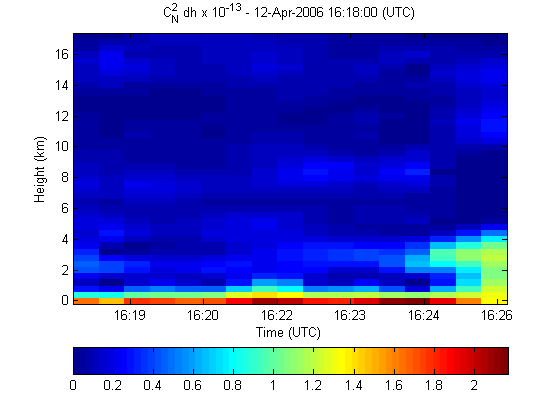}}}
      }
    \mbox{
      \subfigure[]{\scalebox{1.0}{\includegraphics[width=0.45\textwidth, bb=0 0 560 420]{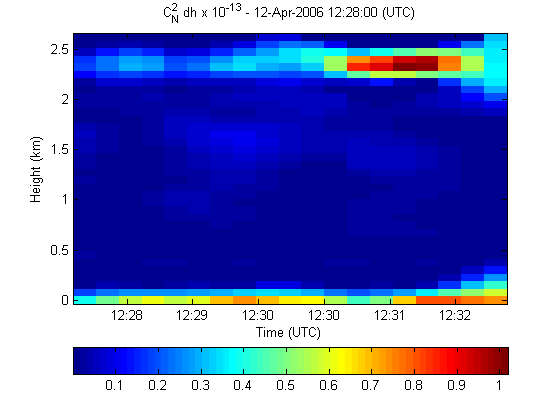}}} \quad
      \subfigure[]{\scalebox{1.0}{\includegraphics[width=0.45\textwidth, bb=0 0 560 420]{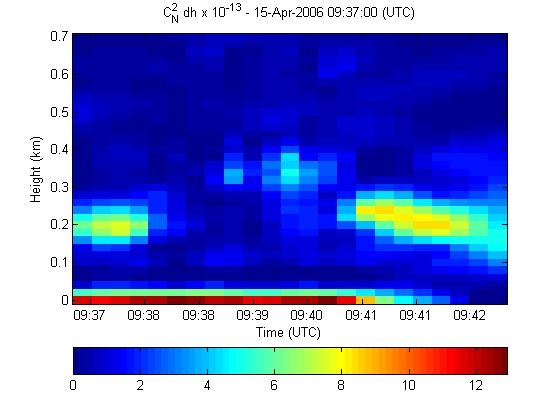}}}
      }
    \caption{SSO Run 7: Examples of consecutive SLODAR turbulence profiles from difference spaced double star targets (height ranges). Each temporal plot represents a group of datasets from the same double star target (similar height sampling) measured during 11-17 April 2006. The vertical axis denotes height (km) and the horizontal axis denotes time (UTC). The color denotes turbulence strength, $C_N^2(h).dh$ ($m^{4-\beta}$). The plots are derived by interpolating the turbulence profiles onto a regular spaced grid at approximately Nyquist sampling. The blank regions represent times having no data. Note, $h$=0~km, defines the height of telescope primary mirror.} \label{sso_grp_profile_run7}
  \end{center}
\end{figure*}

\begin{figure*}[]
  \begin{center}
    \mbox{
      \subfigure[]{\scalebox{1.0}{\includegraphics[width=0.45\textwidth, bb=0 0 603 452]{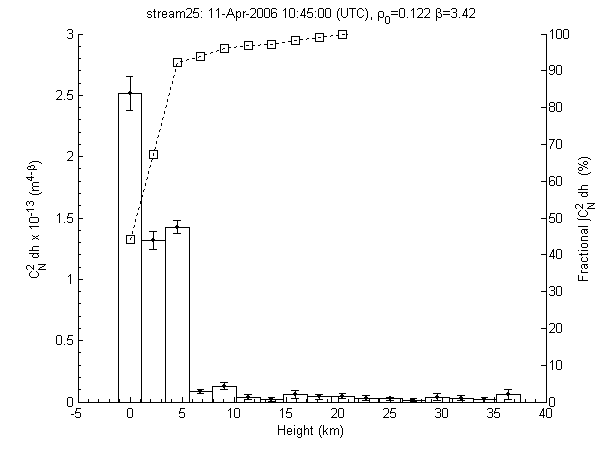}}} \quad
      \subfigure[]{\scalebox{1.0}{\includegraphics[width=0.45\textwidth, bb=0 0 603 452]{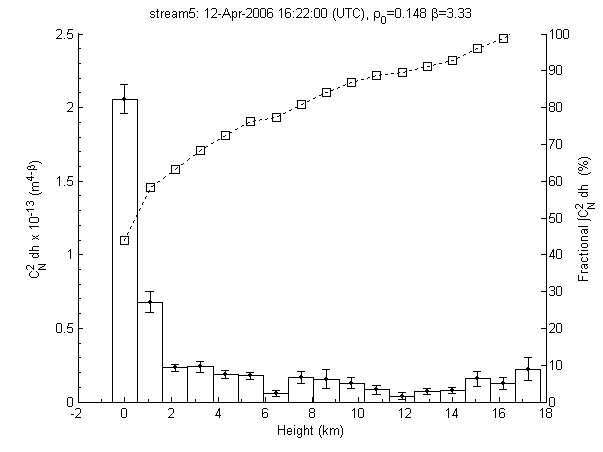}}}
      }
    \mbox{
      \subfigure[]{\scalebox{1.0}{\includegraphics[width=0.45\textwidth, bb=0 0 603 452]{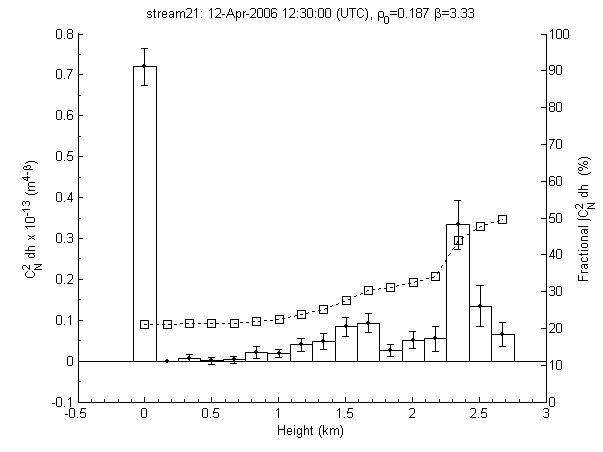}}} \quad
      \subfigure[]{\scalebox{1.0}{\includegraphics[width=0.45\textwidth, bb=0 0 603 452]{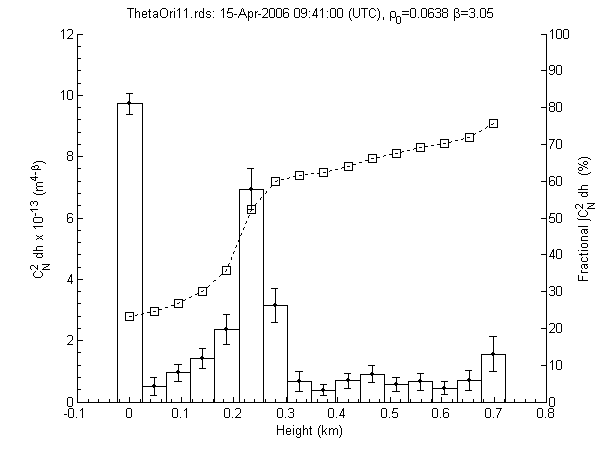}}}
      }
    \caption{SSO Run 7: Examples of individual SLODAR turbulence profiles with each plot representing a single profile as represented in temporal plots of Figure~\ref{sso_grp_profile_run7} measured during 11-17 April 2006. The vertical axis denotes turbulence strength, $C_N^2(h).dh$ ($m^{4-\beta}$), and the horizontal axis denotes height (km). Note, $h$=0~km, defines the height of telescope primary mirror.} \label{sso_indv_profile_run7}
  \end{center}
\end{figure*}

In this section we provide examples of consecutive $C_N^2(h)$ turbulence profiles taken from the same double star target (or `group' datasets having similar time-stamps and height sampling) during the seventh observing run (11-17 April 2006) and  eighth observing run (15-21 June 2006). Note we use the convention that $h$=0~km, defines the height of telescope primary mirror. These observing runs had the highest number of datasets logged (see Table~\ref{tab:slodarrunssso}) and provide a useful visual indicator of the spatial-temporal evolution of the turbulence. Consecutive profiles for the seventh run are plotted in Figure~\ref{sso_grp_profile_run7}. Example individual turbulence profiles from these group datasets are plotted in Figure~\ref{sso_indv_profile_run7}.

The temporal plots of the  turbulence profiles shows a number of dynamical characteristics; (i) intense turbulence occurring near the ground (below 100~m); (ii) turbulent layers that can fluctuate in intensity, appearing in `bursts' with timescales of several minutes. This can have implications for different lines of sight; (iii) appearance to drift in altitude on some occasions rather than disperse before disappearing.

%\subsection{Statistical Parameters}

To quantify the structural distribution of the atmospheric turbulence two statistical parameters are defined, namely $h_{nnnn}$. The $h_{nnnn}$ parameter describes the fractional amount (0.0 to 1.0, with 1.0 being 100\%) of turbulence below $nnnn$~m as measured from the telescope primary mirror. The default value for $nnnn$ is 500~m or $h_{500}$, but due to coarse height resolution sampling, the value of $nnnn$ can be larger, e.g, 750~m or $h_{750}$. The $h_{nnnn}$ allows a qualitative assessment for the performance of ground layer adaptive optics which is favorable if the bulk of the turbulence is near the ground.

The third statistical parameter characterises the power-law slope index of the power spectrum of spatial phase fluctuations, $\beta_{avg}$. The parameter $\beta_{avg}$ is representative of the entire atmospheric turbulence, or averaged contribution of all layer heights. The $\beta_{avg}$ was determined by the best fit of the $\Delta=0$ theoretical covariance impulse response function for $\beta$ ranging from $19/6$ to $23/6$ to the observed auto-covariance function, see~\cite{Butterley2006}. The implications of non-Kolmogorov turbulence ($\beta \neq 11/3$) in the measurement of atmospheric turbulence and astronomical imaging are discussed by ~\cite{Stribling1995} and ~\cite{Goodwin2009}.

 Fits of a non-Kolmogorov exponent to the SLODAR power spectrum (auto-covariance) are not provided in this paper. The fitted data has comparable error bars to~\cite{Butterley2006} and that the best fit within the error bars typically produces exponent values that are non-Kolmogorov  (less than 11/3). An exponent less than 11/3 causes the DIMM seeing to be overestimated due to increased image motion for small apertures~\cite{Goodwin2009}. The power spectrum is relatively insensitive to the outer-scale (Von Karman power spectrum) as noted by ~\cite{Butterley2006}. A fit of the outer-scale would typical need an outer scale smaller than the telescope (1m) which is not typical of other measurements at other observatories (20-40m). The outer-scale fit also has typically larger residuals for the larger offsets in the covariance function compared to the exponent fit.

%\subsection{Turbulence Distribution  ($h_{500}$), Power Law ($\beta_{avg}$) and Seeing FWHM}
%\label{sec:turbulencedistribution}

\begin{figure}[]
  \begin{center}
    \mbox{
      \subfigure[]{\scalebox{1.0}{\includegraphics[width=0.9\columnwidth] {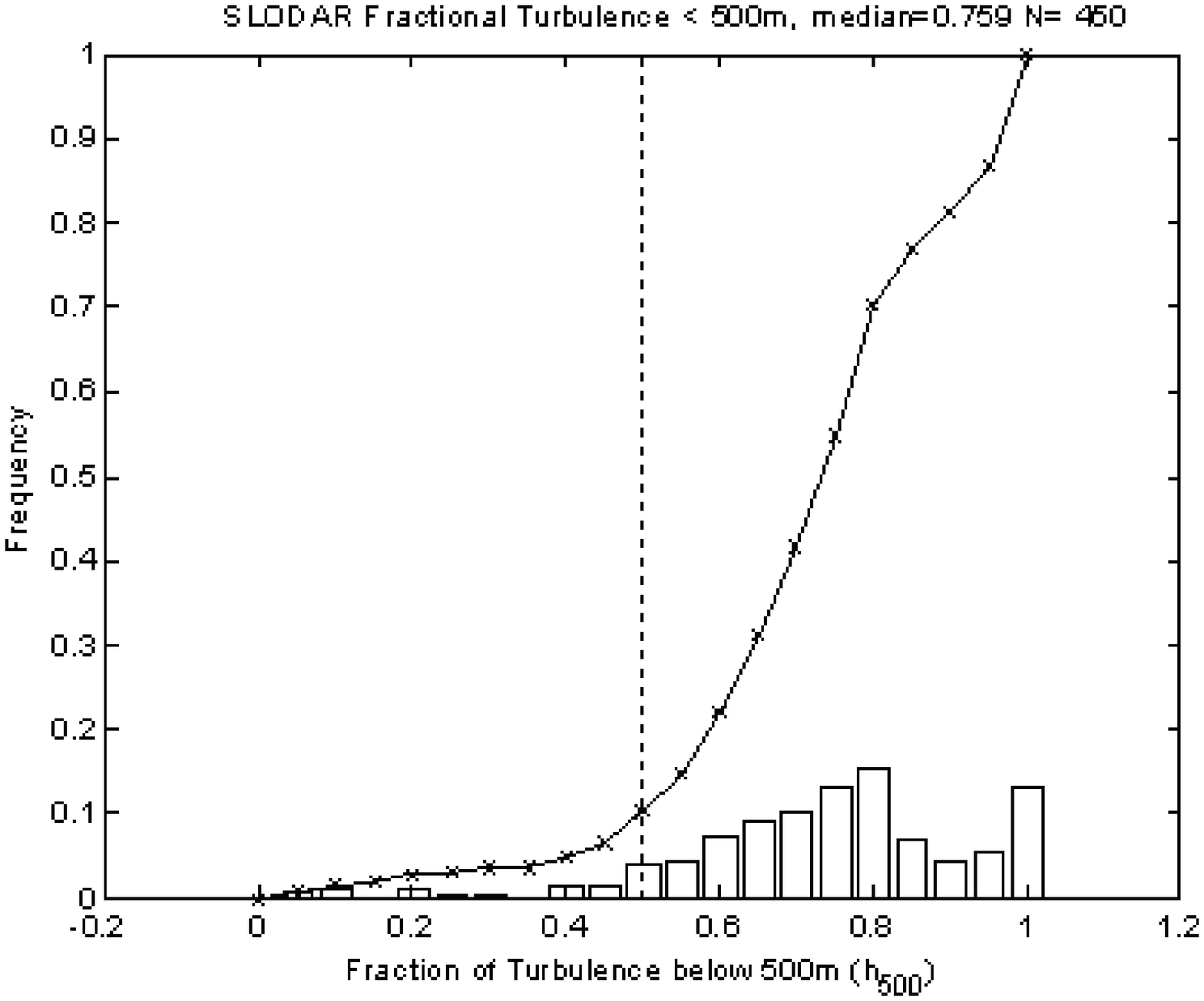}}}
      }
    \mbox{
      \subfigure[]{\scalebox{1.0}{\includegraphics[width=0.9\columnwidth] {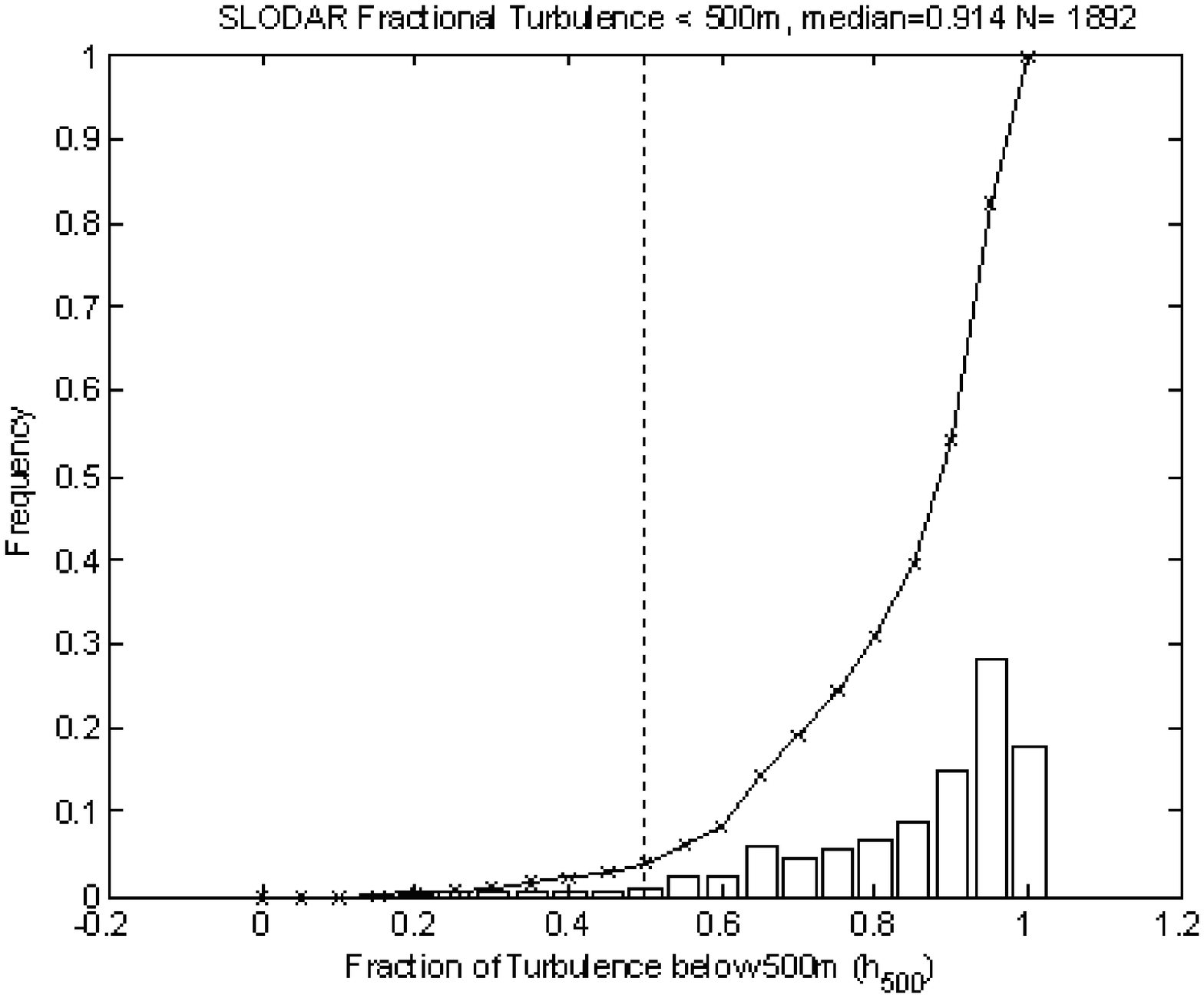}}}
      }
    \caption{Summary plot showing the fraction of turbulence below 500~m based on all observable nights measured during (a) SSO Run 7:  11-17 April 2006 with the median fractional amount of turbulence below 500~m is 76\% (based on 450 datasets); (b) SSO Run 8: 15-21 June 2006 wit the median fractional amount of turbulence below 500~m is 91\% (based on 1892 datasets). } \label{fig:sso_h500Summary_run7}
  \end{center}
\end{figure}

The results measured during 11-17 April 2006 (run 7) and 15-21 June 2006 (run 8)  using the ANU 17$\times$17 SLODAR instrument on the ANU 40" telescope indicate an atmospheric turbulence structure that is dominated by strong ground-layer turbulence. This is evident in the summary $h_{500}$ parameter for all nights as shown in Figure~\ref{fig:sso_h500Summary_run7}. The summary $h_{500}$ parameter indicates that nearly 76\% (run 7) and 91\% (run 8)  of the integrated turbulence is below 500~m. Note that a few  cases of limited height resolution sampling, the $h_{500}$ will be more representative of higher altitudes, e.g.  $h_{750}$, but this is a minority of the datasets.

\begin{figure}[]
  \begin{center}
    \mbox{
      \subfigure[]{\scalebox{1.0}{\includegraphics[width=0.9\columnwidth] {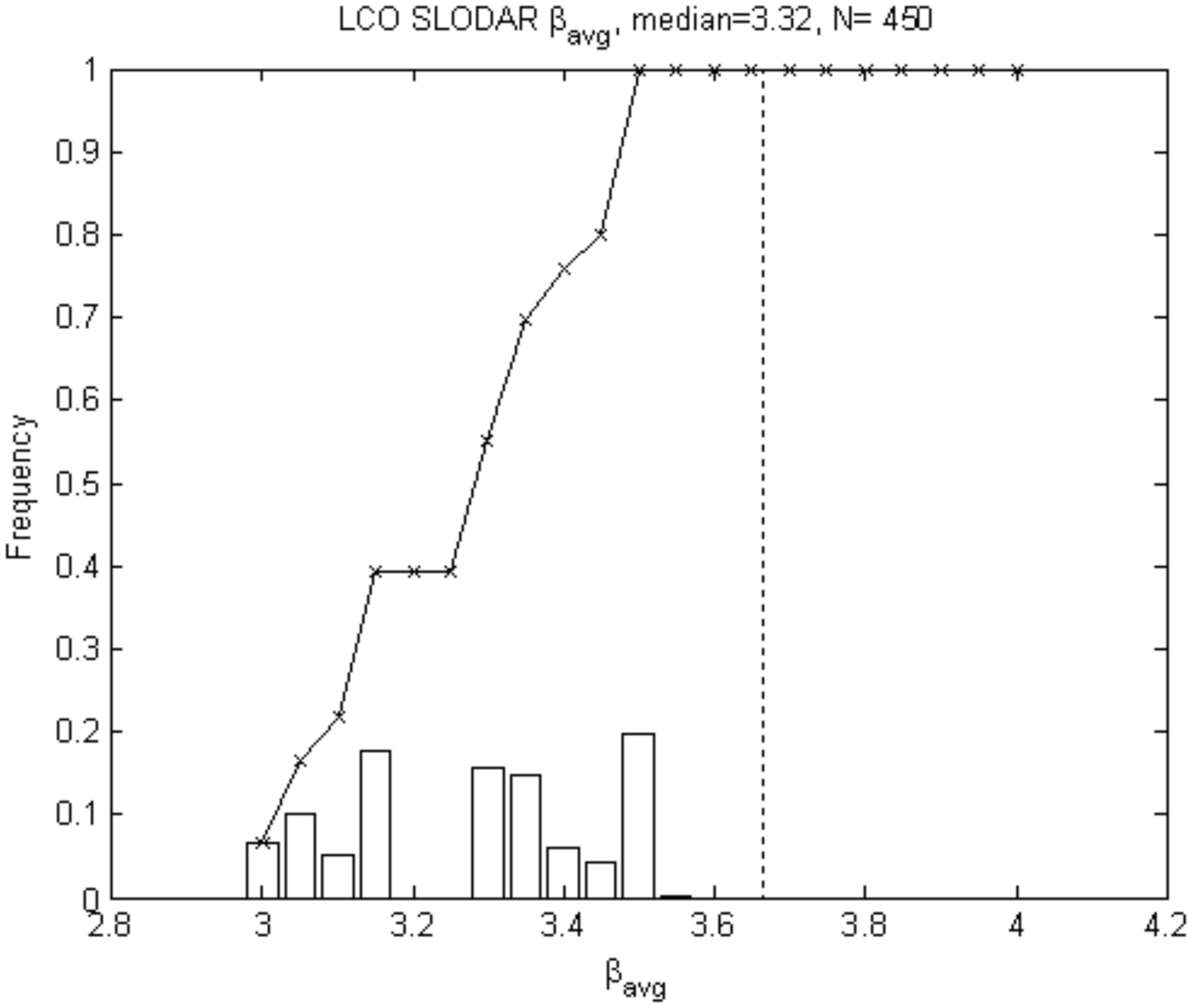}}}
      }
    \mbox{
      \subfigure[]{\scalebox{1.0}{\includegraphics[width=0.9\columnwidth] {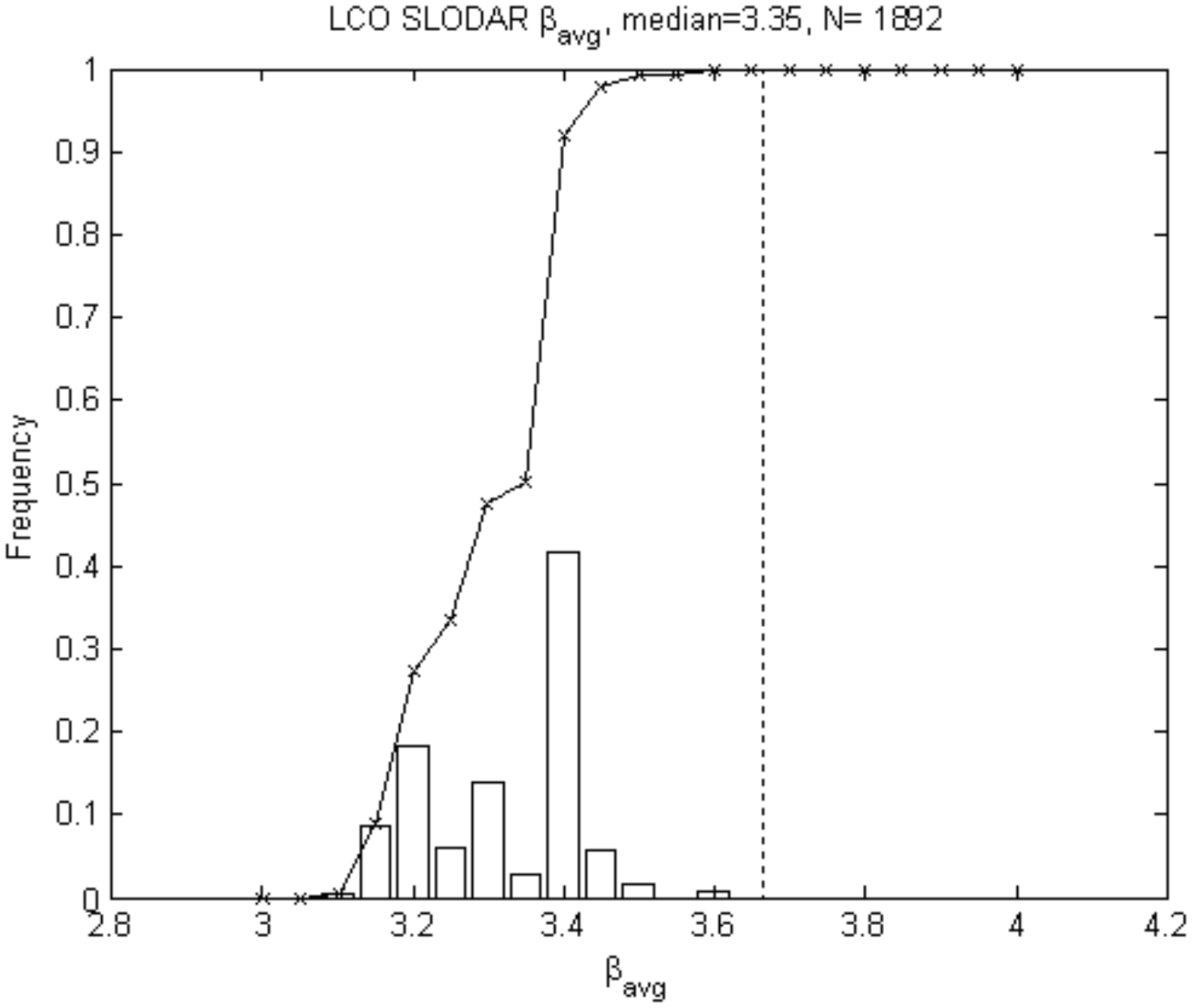}}}
      }
    \caption{Summary plot showing the average power law slope, $\beta_{avg}$, of the spatial power spectrum of phase fluctuations based on all observable nights measured during (a) SSO Run 7: 11-17 April 2006 with the median of 3.32 (based on 450 datasets); (b)  15-21 June 2006 with the median of 3.35 (based on 1892 datasets). For both cases the values are noticeably less than the Kolmogorov value of 3.67 (dashed vertical line).  } \label{fig:sso_BetaAvgSummary_run7}
  \end{center}
\end{figure}

The summary $\beta_{avg}$ parameter for all nights is shown in Figure~\ref{fig:sso_BetaAvgSummary_run7}. The summary $\beta_{avg}$ parameter is found to have a median of 3.32 (run 7) and  3.35 (run 8) as compared to a Kolmogorov value of 3.67. This implies that the strong ground-layer is non-Kolmogorov causing a low $\beta_{avg}$.

\begin{figure}[]
  \begin{center}
    \mbox{
      \subfigure[]{\scalebox{1.0}{\includegraphics[width=0.9\columnwidth] {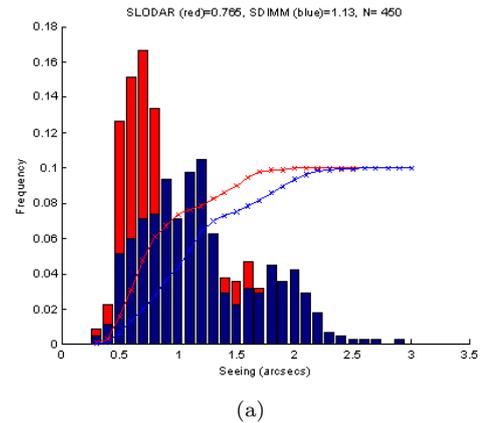}}}
      }
    \mbox{
      \subfigure[]{\scalebox{1.0}{\includegraphics[width=0.9\columnwidth] {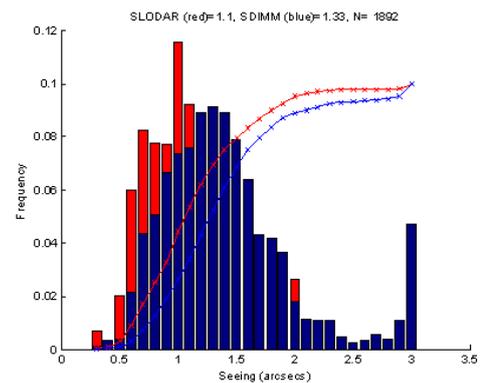}}}
      }
    \caption{Summary plot showing the seeing histograms of the SLODAR (non-Kolmogorov, red) method and DIMM (Kolmogorov, blue) method based on all observable nights measured during (a) SSO Run 7: 11-17 April 2006  (based on 450 datasets); (b)  15-21 June 2006 (based on 1892 datasets). For both cases the seeing valuesare reported at a wavelength of 0.5 microns.  } \label{fig:sso_SeeingSummary_run7}
  \end{center}
\end{figure}

The summary seeing (for a wavelength of 0.5 microns) derived from the SLODAR function-fitting method (non-Kolmogorov analysis) and the DIMM method (Kolmogorov) are shown as a histograms in Figure~\ref{fig:sso_SeeingSummary_run7}. The median seeing values for non-Kolmogorov and DIMM analysis (using SLODAR data) are  0.77" and 1.13"  (run 7) and  1.1" and 1.33"  (run 8). The discrepancy between SLODAR (non-Kolmogorov) and DIMM (Kolmogorov) seeing calculation methods are most likely due to the low $\beta_{avg}$ values in the data (strong non-Kolmogorov effects). It is important to note that the seeing values have the mirror/dome seeing component removed~\citep{Goodwin2007}, which is usually found to be a significant component. We note that the DIMM seeing of the SLODAR data, 1.13" and 1.33" brackets the historical DIMM seeing measurements by \cite{Wood1995} of 1.25".

\subsection{$V(h)$ profiles}

\label{sec:layerwindprofiles}

\begin{figure*}
  \centering
  \includegraphics[width=1.0\textwidth]{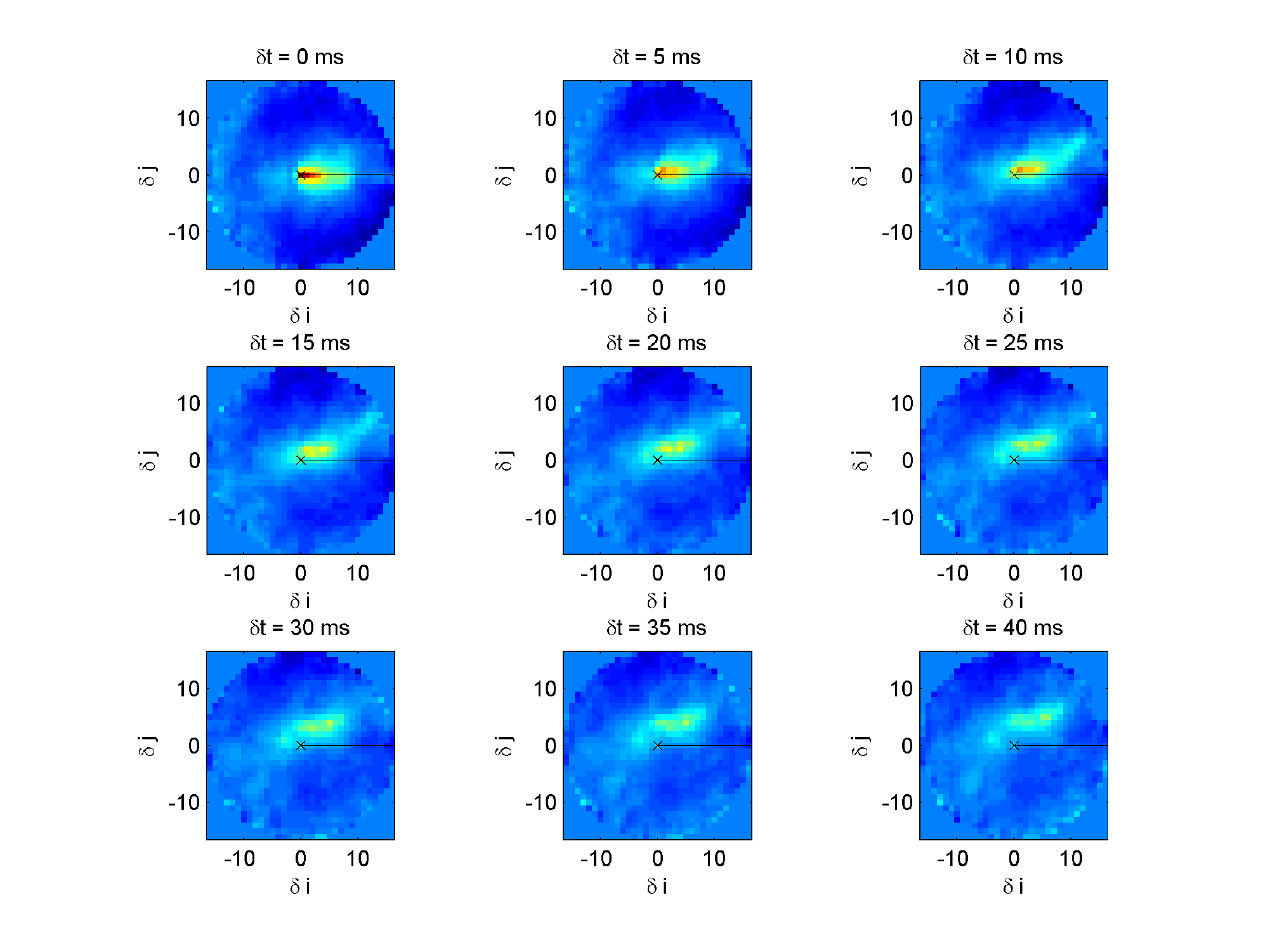}\\
  \caption{SSO Run 7: Example temporal spatial cross-covariance of centroid data for layer wind speed determination as measured during 11-17 April 2006 with the CCD camera (TM6740GE) having a frame rate of 200~fps (area of interest read-out). The example is of the double star $\alpha~Cen$ with height resolution $\delta h$=1.0~km. The double star separation axis (positive heights) is marked with a black line. The temporal offset, $\tau$, is a multiple of the inverse of the frame rate, or 5~ms, starting from top left panel with $\tau$=0~ms and the largest offset located at bottom right panel with $\tau$=40~ms. The pixels represent the sub-aperture offsets ($\delta i$, $\delta j$) with physical size $w$=5.8~cm. The wind speed of a layer can be estimated by $s/\tau$, where $s$ is the physical displacement of covariance peak for a given temporal offset, $\tau$. For this example, four separate layers are detected with speeds (1) 0.5~m/20~ms = 25 m/s (8~km); (2) 0.4~m/40~ms = 10 m/s (4~km); (3) 0.26~m/40~ms = 6.5 m/s (2~km); and (4) 0.18~m/40~ms = 4.5 m/s (0~km). Note, $h$=0~km, defines the height of telescope primary mirror.}\label{fig:sso_wind_run7}
\end{figure*}

The translational velocity information of the turbulent layers can be retrieved by introducing gradual temporal offsets, $\delta t$, in the spatial cross-covariance function of wavefront gradients and observing the corresponding displacement of peaks~\citep{Wilson2002}. The temporal offsets, $\delta t$, are integer multiples of the camera acquisition time (inverse of frame rate), and therefore must be sufficiently short to capture multiple observations of the turbulent layer as it moves across the telescope pupil. Typically, $\delta t < 100~ms$, with camera acquisition times between 5~ms and 50~ms. A turbulence layer moving at velocity, $v$, will have its cross-covariance peak shifted by $v\delta t$ from its location at $\delta t=0$, aligned along the separation axis of the double star. By making several measurements of the spatial shifts in the cross-covariance peak for several sequential temporal offsets, $\delta t$, it is possible to trace the turbulent layer back to the origin to determine both height and velocity information. Examples of a layer wind speed measurement using the temporal spatial cross-covariance of centroid data during the seventh observing run (11-17 April 2006) is shown in Figure~\ref{fig:sso_wind_run7}.

%%%%%%%%%%%%%%%%%%%%

\section{Model-OTP}

A model-OTP is required to summarize the main characteristics of the measured atmospheric turbulence so that adaptive optics simulations can be performed and instrument performance can be predicted. The difficulty is that the atmospheric turbulence
profiles follow a non-stationary process and therefore individual
profiles are not representative for use in adaptive optics simulations.
Of particular interest are the characteristics of the ground-layer and
free-atmospheric turbulence, such as the contribution to the total
turbulence integral (seeing), thickness and intensity of the
ground-layer and if any persistent prominent layers exist.

Model-OTPs to characterise the atmospheric turbulence have been
synthesized from measurements obtained at other astronomical
observatories. \cite{TokovininTravouillon2006} note that previous model-OTPs that are based on the average or median profiles, such as in \cite{Abahamid2004}, do not model the strong variability property of turbulence.
\cite{Abahamid2004} point out that the turbulence intensity at any given altitude changes by several orders of magnitude and that real OTPs are
typically dominated by a few strong layers. Hence the use of only median
or average techniques for OTPs to characterise the atmospheric
turbulence for adaptive optics analysis may be misleading.

An example is the characterisation of the OTP above the Cerro Pachon
(CP) astronomical site. The CP site was initially characterised during
the 1998 Gemini site campaign \citep{Vernin1998} in which seven
discrete layers were modeled by \cite{Ellerbroek2000},
referred to as the ER2000 model. The ER2000 model has been used by
other groups in AO simulations but is insufficient in two respects as noted by \cite{TokovininTravouillon2006}: (i) it does not address the variability
of the real OTP and (ii) it was developed for the needs of classical and
multi-conjugate AO that is mostly affected by high-altitude turbulence.

To resolve the limitations with the ER2000 model, and other similar
models, \cite{TokovininTravouillon2006} propose a new method to
derive a more detailed statistical model-OTP suitable for GLAO analysis
as well as other adaptive optics techniques. The proposed method has already been  used by~\cite{Andersen2006} to model the OTP for Cerro Pachon for wide-field GLAO simulation for Gemini-South. An outline of the methodology for the model-OTP is given by \cite{TokovininTravouillon2006}. Note that the Model-OTP derived in this paper also assumes a Kolmogorov turbulence model.

\subsection{Methodology: Layer Strength Model}

The model-OTP proposed by \cite{TokovininTravouillon2006} separates the ground layer (GL) statistics from the free-atmosphere (FA) statistics
(observed to be independent). The GL zone is defined from the telescope
(10\textendash{}50~m) to some 500~m above the site. The FA zone is defined
as all turbulence above the GL zone.

The \cite{TokovininTravouillon2006} method is well suited to the
SLODAR method. The SLODAR method has the ability to measure the GL and
FA with sufficient height resolution by choosing appropriate double
star separations. The GL can be measured with sufficient height
resolution by selecting double stars having the widest separation.

The OTP is universally defined as the dependence of the refractive
index structure constant $C_N^2$ (measured in units $m^{-2/3}$) at altitude $h$ (measured in units of meters) above sea-level (in this paper, $h$=0~km, defines the height of telescope primary mirror). The turbulence integral, $J$, is defined as:

\begin{equation}
J = \int {C_N^2 \left( h \right)} .dh
\end{equation}

(measured in units $m^{1/3}$) is calculated over some altitude range.

The turbulence integral calculated over the entire height range
covering the atmospheric turbulence (0 to 20~km), the total turbulence
strength, can be expressed as the astronomical `seeing' (units in
arcsecs). The `seeing' is the spot image full width at half-maximum
(FWHM) of an unresolved object (e.g., unresolved star).  The seeing for
observations at a wavelength of $\lambda=500$~nm at the zenith (Kolmogorov turbulence model) is given by:

\begin{equation}
\label{eqn:seeingJ}
\epsilon  = \left[ {{J \mathord{\left/
 {\vphantom {J {\left( {6.8 \times 10^{ - 13} } \right)}}} \right.
 \kern-\nulldelimiterspace} {\left( {6.8 \times 10^{ - 13} } \right)}}} \right]^{0.6}
\end{equation}

As noted by \cite{TokovininTravouillon2006} the seeing is not
additive, hence the preference to use the turbulence integrals, $J$, which
are additive and direct comparisons are possible. The \cite{TokovininTravouillon2006} method calculates the $C_N^2$ turbulence integral, $J$, derived from observations for the ground layer ($J_{GL}$) and the free-atmosphere ($J_{FA}$), then categorizes into `good',
`typical and `bad' conditions. The representative `good' profile is
based on an averaged profile representative of the 1$^{st}$ quartile
(25\%) of $J_{GL}$ and $J_{FA}$, based on averaging observational profiles in the range (15\% to 35\%) to ensure an adequate sample size. Likewise the
`typical' profile based on the range (40\% to 60\%) and the `bad'
profile (65\% to 85\%). This process results in a set of three
representative profiles for each of the GL and FA that have been
averaged separately for each group to reveal typical features.

The \cite{TokovininTravouillon2006} method fits exponential equations
to model the GL intensity and thickness. In this paper, we adopt a
slightly different approach in that prominent layers are modeled as
thin discrete layers, as representative of high-resolution turbulence
profiles with micro-thermal balloon measurements. The thin layers
preserve the relative strengths as well as the turbulence integrals $J_{GL}$
for the GL and $J_{FA}$ for the FA based on the `good' (25\%), `typical'
(50\%) and `bad' (75\%) quartiles.

The $J_{GL}$ was calculated  using the `seeing' parameter (converted to $J$ using Equation~\ref{eqn:seeingJ}) and $h_{500}$ parameter (fractional turbulence below 500~m), using the fact that $J_{GL}=h_{500}J$. The $h_{500}$ parameter can be calculated for all turbulence profiles as the maximum height is greater than 500~m as well as the height resolution of the zero height bin is less than 500~m for almost all observations. Hence the $h_{500}$ parameter is suitable for the calculation of $J_{GL}$. The $J_{FA}$ was calculated using the fact that $J_{FA}=J-J_{GL}$. For the simplicity of calculations, it was assumed that the GL and FA are both modeled by a Kolmogorov turbulence model. With three turbulence profiles based on thin-layers for each of the GL and FA it is possible to construct an OTP model having nine possible outcomes with respective probabilities. 
%The nine model-OTPs are numbered for reference in Table~\ref{tab:optprofilecombinations}.

%\begin{table}[tbp]
%\center
%\begin{tabular}{|l|l|lll|}
%\cline{3-5}
%\multicolumn{1}{c}{} & \multicolumn{1}{c|}{} & \multicolumn{3}{c|}{Ground Layer Model} \\
%\cline{3-5}
%\multicolumn{1}{c}{} & \multicolumn{1}{c|}{} & \multicolumn{1}{c|}{good} & \multicolumn{1}{c|}{typical} & \multicolumn{1}{c|}{bad} \\
%\hline
%\multicolumn{1}{|c|}{Free} & \multicolumn{1}{c|}{good} & \multicolumn{1}{c}{1} & \multicolumn{1}{c}{4} & \multicolumn{1}{c|}{7} \\
%\cline{2-2}
%\multicolumn{1}{|c|}{Atmosphere} & \multicolumn{1}{c|}{typical} & \multicolumn{1}{c}{2} & \multicolumn{1}{c}{5} & \multicolumn{1}{c|}{8} \\
%\cline{2-2}
%\multicolumn{1}{|c|}{Model} & \multicolumn{1}{c|}{bad} & \multicolumn{1}{c}{3} & \multicolumn{1}{c}{6} & \multicolumn{1}{c|}{9} \\
%\hline
%\end{tabular}
%\caption{\label{tab:optprofilecombinations} Possible combinations of GL and FA `good', `typical' and `bad' conditions numbered for referencing in the model-OTP. }
%\end{table}

The models sufficiently cover half of the conditions (25\% to 75\%) expected at the astronomical site within which a probability of 25\% is assigned to `good', 50\% is assigned to `typical' and 25\% is assigned to `bad' for GL and FA turbulence profiles. The total turbulence profile is a combination of the GL and FA turbulence profiles with probability equal to the respective GL and FA probabilities being multiplied.  The assignment of probabilities to all possible turbulence profile outcomes of the model allows a relative importance to be assigned.  The model does not sufficiently represent the extreme cases of very `good' (0\% to 25\%) and very `bad' (75\% to 100\%) conditions. It is noted that the very `bad' condition is of most concern as adaptive optics may not provide sufficient wavefront correction for scientific operations. Sufficient representation for the very `good' and very `bad' conditions can be obtained by extrapolation, scaling the layer relative fractional amounts by the $J_{GL}$ and $J_{FA}$ turbulence integrals, using values from the respective cumulative density function (CDF) plots.

\subsubsection{Ground Layer Model-OTP}

To derive the GL model-OTP we consider three model turbulence profiles derived from averaging a group of observational profiles in intervals centered on the cumulative density function quartiles for the $J_{GL}$ turbulence integral. The `good' model turbulence profile is derived from averaging profiles having maximum sampling height, $H_{max}$, greater than 500~m but less than 2000~m (ground-layer sampling) within the 15\% to 35\% interval of the cumulative values of $J_{GL}$. Likewise for the `typical' model turbulence profile, within the 40\% to 60\% interval, and `bad' model turbulence profile, within the 65\% to 85\% interval. The relatively small value of $N_{OTP}$ for GL analysis reflects the fact that only a limited number of datasets suitable for GL sampling at SSO were observed. The values relating to the GL model turbulence profiles are listed in Table~\ref{tab:modelotp_ssoall_gl_cumlevels}.

\begin{table*}[tbp]
\begin{center}
\caption{ Levels of the cumulative distributions of $J_{GL}$ used in the calculation of a representative ground layer profiles, `good', `typical' and `bad' for SSO (Run 1-8: May 2005 to June 2006).  }\label{tab:modelotp_ssoall_gl_cumlevels}
\begin{tabular}{llllllll}

\hline   & lower & upper & $N_{OTP}$  &  $\overline{H_{max}}$  &  $\overline{\delta h}$  &  $\overline{J_{GL}}$ & $\overline{\epsilon_{GL}}$ \\
\hline Good & 15\% & 35\% & 17 & 1.0728 & 0.0815     & 5.1866 & 0.8483 \\
Typical & 40\% & 60\% & 10 & 0.7211 & 0.1083 & 8.7999 & 1.1662 \\
Bad & 65\% & 85\% & 32 & 0.7985 & 0.0722    & 13.5809   & 1.5124 \\
\hline
\end{tabular}
\end{center}
\end{table*}

From Table~\ref{tab:modelotp_ssoall_gl_cumlevels} we see that some intervals have more turbulence profiles, due to the $J_{GL}$ distribution consisting of a mixture of both GL and FA turbulence profiles, with a portion not meeting the maximum sampling height criterion. The average maximum height range, $\overline{H_{max}}$, is 0.72~km to 1.07~km and the average height resolution, $\overline{\delta h}$, is 72~m to 108~m.

\begin{figure*}[]
  \begin{center}
    \mbox{
      \subfigure[]{\scalebox{1.0}{\includegraphics[width=0.45\textwidth,bb=0 0 560 420]{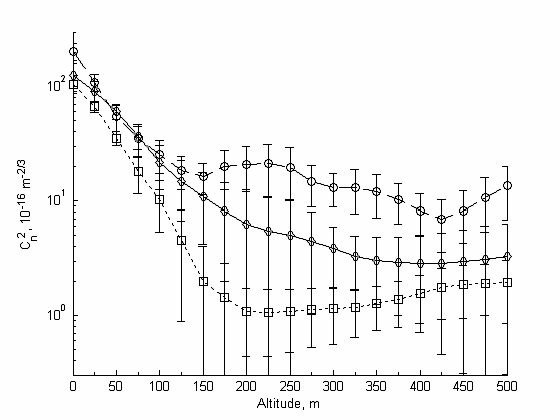}}} \quad
      \subfigure[]{\scalebox{1.0}{\includegraphics[width=0.45\textwidth, bb=0 0 560 420 ]{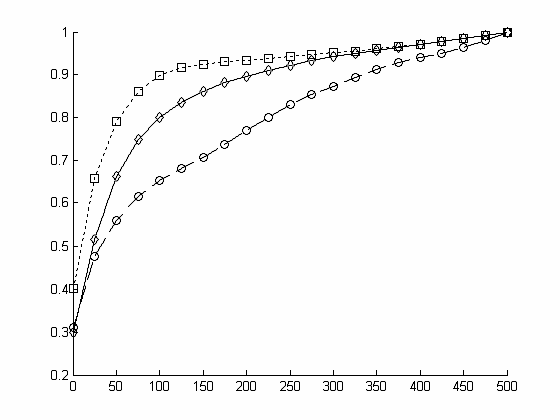}}}
      }
    \caption[Ground-Layer model-OTP and CDF for the SSO (Run 1-8: May 2005 to June 2006) runs.]{Continuous GL model-OTP for SSO (Run 1-8: May 2005 to June 2006): squares `good'; diamonds `typical' and circles `bad' (a) averaged profiles, error bars are 95\% confidence interval and (b) corresponding CDF profiles. }
    \label{fig:modelotp_ssoall_gl_modelotp}
  \end{center}
\end{figure*}

Figure~\ref{fig:modelotp_ssoall_gl_modelotp} shows the GL model turbulence profiles obtained from averaging GL profiles within certain interval ranges of the $J_{GL}$ distribution to represent `good', `typical' and `bad' seeing conditions, as summarized in Table~\ref{tab:modelotp_ssoall_gl_cumlevels}.

A thin-layer GL model-OTP is required to be derived from the continuous GL model-OTP. A thin-layer GL model-OTP is an accurate representation of high resolution OTPs~\citep{Azouit2005}, as well as compatible for adaptive optics simulations (using phase screens to represent thin layers). For the thin-layer GL model-OTP we define two layers at heights 37.5~m and 250~m to best represent the continuous GL model-OTP, Figure~\ref{fig:modelotp_ssoall_gl_modelotp}. To calculate the strengths of the layers an integral (lower,upper) centered on the layer heights of the continuous GL model-OTP is performed. To compare how well layer integrals explain the model, the total layer integral (37.5~m + 250~m) is compared with the total GL integral from 0~m to 500~m (calculated from the continuous GL model-OTP). The results are listed in Table~\ref{tab:modelotp_ssoall_gl_modellayerint}.

\begin{table*}[tbp]
\center
\begin{tabular}{|l|ll|lll|}
\cline{2-6}
\multicolumn{1}{c|}{} & \multicolumn{2}{c|}{Range (m)} & \multicolumn{3}{c|}{Integral, $J$} \\
\hline
\multicolumn{1}{|c|}{Layer Height (m)} & \multicolumn{1}{c|}{lower} & \multicolumn{1}{c|}{upper} & \multicolumn{1}{c|}{good} & \multicolumn{1}{c|}{typical} & \multicolumn{1}{c|}{bad} \\
\hline
\multicolumn{1}{|c|}{37.5} & \multicolumn{1}{c}{0} & \multicolumn{1}{c|}{150} & \multicolumn{1}{c}{4.6104} & \multicolumn{1}{c}{6.9491} & \multicolumn{1}{c|}{8.3327} \\
\multicolumn{1}{|c|}{250} & \multicolumn{1}{c}{150} & \multicolumn{1}{c|}{350} & \multicolumn{1}{c}{0.2431} & \multicolumn{1}{c}{1.0870} & \multicolumn{1}{c|}{3.3985} \\
\hline
\multicolumn{1}{|c|}{Total (37.5+250)} & \multicolumn{1}{c}{} & \multicolumn{1}{c|}{} & \multicolumn{1}{c}{4.8535} & \multicolumn{1}{c}{8.0361} & \multicolumn{1}{c|}{11.7312} \\
\hline
\multicolumn{1}{|c|}{} & \multicolumn{1}{c}{} & \multicolumn{1}{c|}{} & \multicolumn{1}{c}{} & \multicolumn{1}{c}{} & \multicolumn{1}{c|}{} \\
\hline
\multicolumn{1}{|c|}{Integral  (0-500)} & \multicolumn{1}{c}{0} & \multicolumn{1}{c|}{500} & \multicolumn{1}{c}{5.1866} & \multicolumn{1}{c}{8.7999} & \multicolumn{1}{c|}{13.5809} \\
\hline
\multicolumn{1}{|c|}{\% Explained} & \multicolumn{1}{c}{} & \multicolumn{1}{c|}{} & \multicolumn{1}{c}{93.5782} & \multicolumn{1}{c}{91.3203} & \multicolumn{1}{c|}{86.3805} \\
\hline
\end{tabular}
\caption{\label{tab:modelotp_ssoall_gl_modellayerint} Turbulence integrals for the thin-layer model-OTP for the ground layer, $J$ in units of $10^{-13} m^{1/3}$ for SSO (Run 1-8: May 2005 to June 2006). }
\end{table*}

%The relative fractional strengths of the model thin layers are listed in Table~\ref{tab:modelotp_ssoall_gl_modellayerfrac}.
%
%\begin{table}[tbp]
%\center
%\begin{tabular}{|l|lll|}
%\cline{2-4}
%\multicolumn{1}{l|}{} & \multicolumn{3}{c|}{Turbulence Fraction} \\
%\hline
%Layer Height (m) & \multicolumn{1}{l|}{good} & \multicolumn{1}{l|}{typical} & bad \\
%\hline
%37.5 & 0.9499     & 0.8647     & 0.7103     \\
%250  & 0.0501     & 0.3124     & 0.2897     \\
%\hline
%Total (37.5+250) & 1.0000 & 1.0000 & 1.0000 \\
%\hline
%\end{tabular}
%\caption{\label{tab:modelotp_ssoall_gl_modellayerfrac} Turbulence fractional strengths for the thin-layer model-OTP representing the ground layer for SSO (Run 1-8: May 2005 to June 2006).}
%\end{table}

The fractional strengths can be multiplied by $J_{GL}$ according to cumulative level values obtained from its corresponding CDF plot. The model is based on the cumulative levels 25\%, 50\% and 75\%, but other levels can be used for extreme conditions, eg 10\%, 50\% and 90\%, providing some flexibility for a custom model-OTP.

The turbulence integral strengths of the model thin layers are calculated by multiplying the fractional amounts with the cumulative levels 25\%, 50\% and 75\% of $J_{GL}$. The final model thin layers and their turbulence integral, with total ground layer seeing, $\epsilon_{GL}$, and model probability, are listed in Table~\ref{tab:modelotp_ssoall_gl_modellayerint_final}.

\begin{table}[tbp]
\center
\begin{tabular}{|l|lll|}
\cline{2-4}
\multicolumn{1}{l|}{} & \multicolumn{3}{c|}{Model Integral, $J$} \\
\hline
Layer Height (m) & \multicolumn{1}{l|}{good} & \multicolumn{1}{l|}{typical} & bad \\
\hline
37.5 & 4.7129 & 7.3989 & 9.4363 \\
250 & 0.2485 & 1.1574 & 3.8486 \\
\hline
Total (37.5+250) & 4.9614 & 8.5562 & 13.2849 \\
\hline
$\epsilon_{GL}$ & 0.8277  & 1.1478 & 1.4945  \\
\hline
Probability & 25\% & 50\% & 25\% \\
\hline
\end{tabular}
\caption{\label{tab:modelotp_ssoall_gl_modellayerint_final} Final turbulence integrals for the GL thin-layer model-OTP, $J$ in units of $10^{-13} m^{1/3}$ for SSO (Run 1-8: May 2005 to June 2006).}
\end{table}

\subsubsection{Free Atmosphere Model-OTP}
\label{sec:freeatmosphremodelotp_ssoall}
To derive the FA model we follow a similar approach to that described for the GL model. This involves deriving three model turbulence profiles from averaging a group of observational profiles within representative intervals of cumulative density function values for the $J_{FA}$ turbulence integral. The `good' model turbulence profile is derived from averaging profiles having maximum sampling height, $H_{max}$, greater than 16000~m but less than 20000~m (free-atmosphere sampling) within the 15\% to 35\% interval of the cumulative values of $J_{FA}$. Likewise for the `typical' model turbulence profile, within the 40\% to 60\% interval, and `bad' model turbulence profile, within the 65\% to 85\% interval. The values relating to the FA model turbulence profiles are listed in Table~\ref{tab:modelotp_ssoall_fa_cumlevels}.

\begin{table*}[tbp]
\begin{center}
\caption{ Levels of the cumulative distributions of $J_{FA}$ used in the calculation of a representative ground layer profiles, `good', `typical' and `bad' for SSO (Run 1-8: May 2005 to June 2006).  }\label{tab:modelotp_ssoall_fa_cumlevels}
\begin{tabular}{llllllll}

\hline   & lower & upper & $N_{OTP}$  &  $\overline{H_{max}}$  &  $\overline{\delta h}$  &  $\overline{J_{GL}}$ & $\overline{\epsilon_{GL}}$ \\
\hline Good & 15\% & 35\% & 147 & 17.0076 & 1.0951 & 0.3702     & 0.1719 \\
Typical & 40\% & 60\% & 118 & 17.1918 & 1.3382    & 1.1983    & 0.3509 \\
Bad & 65\% & 85\% & 58 & 17.3392     & 1.1926     & 2.5276     & 0.5503 \\
\hline
\end{tabular}
\end{center}
\end{table*}

From Table~\ref{tab:modelotp_ssoall_fa_cumlevels} we see that some intervals have more turbulence profiles, due to the $J_{FA}$ distribution consisting of a mixture of both GL and FA turbulence profiles, with a portion not meeting the maximum sampling height criteria. The average maximum height range, $\overline{H_{max}}$, is 17.0~km to 17.3~km and the average height resolution, $\overline{\delta h}$, is 1095~m to 1338~m.

\begin{figure*}[htbp]
  \begin{center}
    \mbox{
      \subfigure[]{\scalebox{1.0}{\includegraphics[width=0.45\textwidth,bb=0 0 560 420]{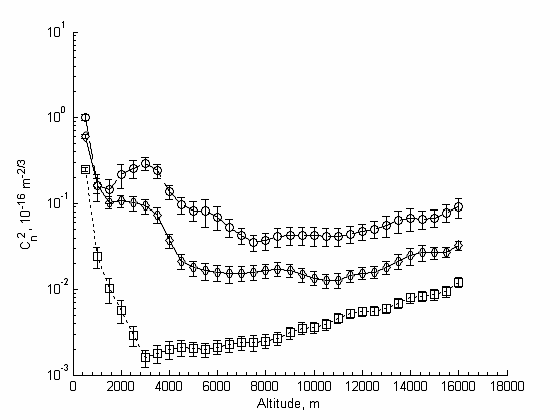}}} \quad
      \subfigure[]{\scalebox{1.0}{\includegraphics[width=0.45\textwidth, bb=0 0 560 420 ]{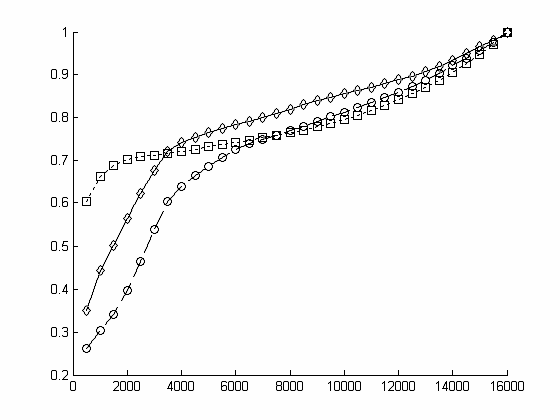}}}
      }
    \caption[Free atmosphere model-OTP and CDF for the SSO (Run 1-8: May 2005 to June 2006) runs.]{Continuous FA model-OTP for SSO (Run 1-8: May 2005 to June 2006): squares `good'; diamonds `typical' and circles `bad' (a) averaged profiles, error bars are 95\% confidence interval (b) corresponding CDF profiles. }
    \label{fig:modelotp_ssoall_fa_modelotp}
  \end{center}
\end{figure*}

Figure~\ref{fig:modelotp_ssoall_fa_modelotp} shows the FA model turbulence profiles obtained from averaging FA profiles within certain interval ranges of the $J_{FA}$ distribution to represent `good', `typical' and `bad' seeing conditions, as summarized in Table~\ref{tab:modelotp_ssoall_fa_cumlevels}.

A thin-layer FA model-OTP is required to be derived from the continuous FA model-OTP. A thin-layer FA model-OTP is an accurate representation of high resolution OTPs~\citep{Azouit2005} as well as compatible for adaptive optic simulations (using phase screens to represent thin layers). For the thin-layer FA model-OTP we define five layers at heights 1000~m, 3000~m, 6000~m, 9000~m and 13500~m to best represent the continuous FA model-OTP, Figure~\ref{fig:modelotp_ssoall_fa_modelotp}. To calculate the strengths of the layers an integral (lower,upper) centered on the layer heights of the continuous FA model-OTP is performed. To compare how well layer integrals explain the model, the total layer integral (1000~m+3000~m+6000~m+9000~m+13500~m) is compared with the total FA integral from 500~m to 16000~m (calculated from the continuous FA model-OTP). The results are listed in Table~\ref{tab:modelotp_ssoall_fa_modellayerint}.

The FA model-OTP in Table~\ref{tab:modelotp_ssoall_fa_modellayerint} has a gap in binning between 10.5km and 12km as well above 15km. This is because we rarely saw any evidence for turbulence at those heights in the continuous FA profile of Figure~\ref{fig:modelotp_ssoall_fa_modelotp}. There was also an attempt to keep the number of model layers to minimum as well as keeping equal bin widths of 3000m. The gap in binning will result in a slight bias to increase the relative strength of the FA lower layers causing the isoplanatic angle to be slightly larger (providing an optimistic estimate). The value of the bias is expected to be small.

\begin{table*}[tbp]
\center
\begin{tabular}{|l|ll|lll|}
\cline{2-6}
\multicolumn{1}{c|}{} & \multicolumn{2}{c|}{Range (m)} & \multicolumn{3}{c|}{Integral, $J$} \\
\hline
\multicolumn{1}{|c|}{Layer Height (m)} & \multicolumn{1}{c|}{lower} & \multicolumn{1}{c|}{upper} & \multicolumn{1}{c|}{good} & \multicolumn{1}{c|}{typical} & \multicolumn{1}{c|}{bad} \\
\hline
\multicolumn{1}{|c|}{1000} & \multicolumn{1}{c}{500} & \multicolumn{1}{c|}{1500} & \multicolumn{1}{c}{0.0770} & \multicolumn{1}{c}{0.2577} & \multicolumn{1}{c|}{0.3675} \\
\multicolumn{1}{|c|}{3000} & \multicolumn{1}{c}{1500} & \multicolumn{1}{c|}{4500} & \multicolumn{1}{c}{0.0100} & \multicolumn{1}{c}{0.2358} & \multicolumn{1}{c|}{0.6265} \\
\multicolumn{1}{|c|}{6000} & \multicolumn{1}{c}{4500} & \multicolumn{1}{c|}{7500} & \multicolumn{1}{c}{0.0065} & \multicolumn{1}{c}{0.0499} & \multicolumn{1}{c|}{0.1974} \\
\multicolumn{1}{|c|}{9000} & \multicolumn{1}{c}{7500} & \multicolumn{1}{c|}{10500} & \multicolumn{1}{c}{0.0093} & \multicolumn{1}{c}{0.0466} & \multicolumn{1}{c|}{0.1222} \\
\multicolumn{1}{|c|}{13500} & \multicolumn{1}{c}{12000} & \multicolumn{1}{c|}{15000} & \multicolumn{1}{c}{0.0209} & \multicolumn{1}{c}{0.0641} & \multicolumn{1}{c|}{0.1810} \\
\hline
\multicolumn{1}{|c|}{Total (1000+3000+6000+9000+13500)} & \multicolumn{1}{c}{} & \multicolumn{1}{c|}{} & \multicolumn{1}{c}{0.1237} & \multicolumn{1}{c}{0.6541} & \multicolumn{1}{c|}{1.4946} \\
\hline
\multicolumn{1}{|c|}{} & \multicolumn{1}{c}{} & \multicolumn{1}{c|}{} & \multicolumn{1}{c}{} & \multicolumn{1}{c}{} & \multicolumn{1}{c|}{} \\
\hline
\multicolumn{1}{|c|}{Integral  (500-16000)} & \multicolumn{1}{c}{500} & \multicolumn{1}{c|}{16000} & \multicolumn{1}{c}{0.1409} & \multicolumn{1}{c}{0.7032} & \multicolumn{1}{c|}{1.6387} \\
\hline
\multicolumn{1}{|c|}{\% Explained} & \multicolumn{1}{c}{} & \multicolumn{1}{c|}{} & \multicolumn{1}{c}{87.7795} & \multicolumn{1}{c}{93.0185} & \multicolumn{1}{c|}{91.2053} \\
\hline
\end{tabular}
\caption{\label{tab:modelotp_ssoall_fa_modellayerint} Turbulence integrals for the thin-layer model-OTP for the free atmosphere, $J$ in units of $10^{-13} m^{1/3}$ for SSO (Run 1-8: May 2005 to June 2006). }
\end{table*}

%The relative fractional strengths of the model thin layers are listed in Table~\ref{tab:modelotp_ssoall_fa_modellayerfrac}.
%
%\begin{table*}[tbp]
%\center
%\begin{tabular}{|l|lll|}
%\cline{2-4}
%\multicolumn{1}{l|}{} & \multicolumn{3}{c|}{Turbulence Fraction} \\
%\hline
%Layer Height (m) & \multicolumn{1}{l|}{good} & \multicolumn{1}{l|}{typical} & bad \\
%\hline
%1000 & 0.6220 & 0.3939 & 0.2459 \\
%3000 & 0.0812 & 0.3605 & 0.4192 \\
%6000 & 0.0525 & 0.0763 & 0.1321 \\
%9000 & 0.0750 & 0.0713 & 0.0818 \\
%13500 & 0.1692 & 0.0980 & 0.1211 \\
%\hline
%Total (3000+6000+9000+13500) & 1.0000 & 1.0000 & 1.0000 \\
%\hline
%\end{tabular}
%\caption{\label{tab:modelotp_ssoall_fa_modellayerfrac} Turbulence fractional strengths for the thin-layer model-OTP representing the free atmosphere for SSO (Run 1-8: May 2005 to June 2006).}
%\end{table*}

The fractional strengths can be multiplied by $J_{FA}$ according to cumulative levels obtained from its CDF plot. The model is based on the cumulative levels 25\%, 50\% and 75\%, but for other levels extreme conditions can be used, e.g. 10\%, 50\% and 90\%, providing some flexibility for a custom model-OTP.

The turbulence integral strengths of the model thin layers are calculated by multiplying the fractional amounts with the cumulative levels 25\%, 50\% and 75\% of $J_{FA}$. The final model thin layers and their turbulence integral, with the total free atmosphere seeing, $\epsilon_{FA}$, and model probability, are listed in Table~\ref{tab:modelotp_ssoall_fa_modellayerint_final}.

The total integrated turbulence of the thin-layer FA model as shown in Table~\ref{tab:modelotp_ssoall_fa_modellayerint_final} is approximately twice that of the total integrated turbulence of binned layer continuous FA model as shown in Table~\ref{tab:modelotp_ssoall_fa_modellayerint}. A likely explanation of the scale factor of $\sim2$  may be the result that the parameter $J_{FA}$ has been overestimated (turbulence integral in the free-atmosphere computed based on total seeing and $H_{500}$) compared with the direct turbulence integrals of the averaged measured profiles. The total strength of the weaker free-atmosphere layers (not location) as measured by SLODAR is somewhat underestimated (low S/N). The factor of $\sim2$ scaling increase makes the free-atmosphere seeing in the final model a conservative estimate.

\begin{table*}[tbp]
\center
\begin{tabular}{|l|lll|}
\cline{2-4}
\multicolumn{1}{l|}{} & \multicolumn{3}{c|}{Model Integral, $J$} \\
\hline
Layer Height (m) & \multicolumn{1}{l|}{good} & \multicolumn{1}{l|}{typical} & bad \\
\hline
1000 & 0.2413 & 0.4772 & 0.6677 \\
3000 & 0.0315 & 0.4368 & 1.1384 \\
6000 & 0.0204 & 0.0924 & 0.3588 \\
9000 & 0.0291 & 0.0863 & 0.2221 \\
13500 & 0.0656 & 0.1187 & 0.3289 \\
\hline
Total (3000+6000+9000+13500) & 0.3880 & 1.2115 & 2.7159 \\
\hline
$\epsilon_{FA}$ & 0.1794 & 0.3552 & 0.5766 \\
\hline
Probability & 25\% & 50\% & 25\% \\
\hline
\end{tabular}
\caption{\label{tab:modelotp_ssoall_fa_modellayerint_final} Final turbulence integrals for the FA thin-layer model-OTP, $J$ in units of $10^{-13} m^{1/3}$ for SSO (Run 1-8: May 2005 to June 2006).}
\end{table*}

\subsection{Methodology: Layer Wind Speed and Direction Model}

To model the turbulent layer speeds a Bufton wind speed profile model is used (see Equation~\ref{eqn:buftonwindsitechar}). To model the `good', `typical' and `bad' conditions, three separate Bufton wind profiles are presented based on data found in literature ~\citep{Azouit2005, Tokovinin2003b, Avila2003}. The wind profiles are not based on fits to our data as it is not always possible to obtain a sufficient sample of  layer wind speeds with our SLODAR data due to (i) low camera frame rates; (ii) finite camera exposures; (iii) weakness of a layer; and (iv) short boiling lifetimes (layer de-correlates rapidly). The Bufton wind profile model is given by

\begin{equation}
\label{eqn:buftonwindsitechar}
v(h) = v_G+v_T exp \left[-\left(\frac{h-H_T}{L_T}\right)^2\right]
\end{equation}

where $v_G$ denotes the wind velocity at low altitude, $v_T$ denotes the wind velocity at the tropopause, $H_T$ denotes the height of tropopause, $L_T$ denotes the thickness of the tropopause layer.

\begin{table*}[tbp]
\begin{center}
\caption{Bufton wind model. }  \label{tab:buftonmodel}
\begin{tabular}{lllll}
\hline {} & {$v_G$} & {$v_T$} & {$H_T$} &  {$L_T$} \\
\hline Good & 2 & 30 & 9000 & 4000 \\
Typical & 5 & 35 & 10000 & 5000 \\
Bad & 8 & 40 & 11000 & 6000 \\
\hline
\end{tabular}
\end{center}
\end{table*}

The parameters of the Bufton wind model for `good', `typical' and `bad' conditions used in the analysis are listed in Table~\ref{tab:buftonmodel}. The wind model associates strong ground layer wind speeds with stronger free-atmosphere wind speeds that have a broader upper atmosphere extent ~\citep{Azouit2005, Tokovinin2003b, Avila2003}.  To simplify the model, the `good', `typical' and `bad' conditions are represented by the same profile and referenced to the ground-layer wind direction, set to 0 degrees. The free atmosphere layers can travel in a direction perpendicular to the ground-layer direction. The wind direction model, $\psi(h)$ for `good', `typical' and `bad' conditions is defined as: 

\begin{equation}
\label{eqn:winddirection}
\psi(h) = a(1-exp(-h/b))
\end{equation}

where $a$ = 101.93 and $b$ = 3255.4 are used.

%\begin{figure*}[]
%  \begin{center}
%    \mbox{
%      \subfigure[]{\scalebox{1.0}{\includegraphics[width=0.45\textwidth,bb=0 0 560 420]{modelotp_windspeed.png}}} \quad
%      \subfigure[]{\scalebox{1.0}{\includegraphics[width=0.45\textwidth, bb=0 0 560 420 ]{modelotp_winddirection.png}}}
%      }
%    \caption[Plots of the wind model for model-OTP]{Plots of the wind model for conditions `good' (dotted), `typical' (solid) and `bad' (dashed) for SSO and LCO model-OTP (a) wind speed and (b) wind direction. }
%    \label{fig:modelotpwindmodel}
%  \end{center}
%\end{figure*}

\subsubsection{Turbulent Layer Wind Model}
\label{sec:turblayerwindmodel_ssoall}

\begin{figure*}[htbp]
  \begin{center}
    \mbox{
      \subfigure[]{\scalebox{1.0}{\includegraphics[width=0.45\textwidth,bb=0 0 560 420]{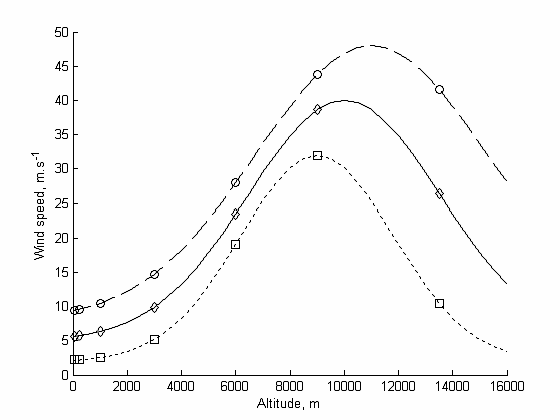}}} \quad
      \subfigure[]{\scalebox{1.0}{\includegraphics[width=0.45\textwidth, bb=0 0 560 420 ]{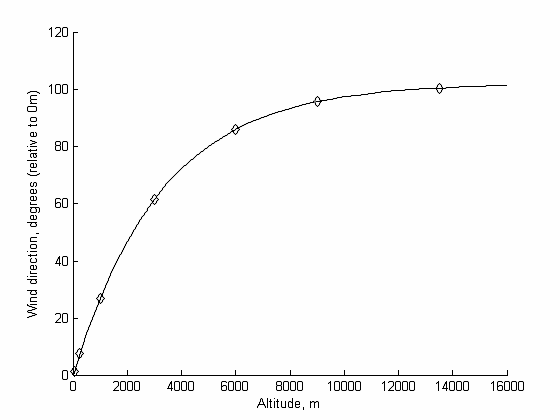}}}
      }
    \caption[Free atmosphere model-OTP and CDF for the SSO (Run 1-8: May 2005 to June 2006) runs.]{Wind model-OTP for the SSO (Run 1-8: May 2005 to June 2006) runs: squares `good'; diamonds `typical' and circles `bad' (a) Bufton wind profile (b) Wind direction (empirical). Layer heights for model-OTP for SSO (Run 1-8: May 2005 to June 2006) model are marked as symbols.}
    \label{fig:modelotp_ssoall_windmodel}
  \end{center}
\end{figure*}

To model the turbulent layer wind speed and direction for the SSO (Run 1-8: May 2005 to June 2006) runs a Bufton wind speed model together with layer height information from the model-OTP is used. The model-OTP layer heights for the ground-layer and free-atmosphere  are used in the Bufton equation to obtain the model layer wind speed values. The model relative layer wind directions are obtained by using the model-OTP layer heights in the wind direction, $\phi(h)$, defined in Equation~\ref{eqn:winddirection}. The wind speed model and wind direction model values for the SSO (Run 1-8: May 2005 to June 2006) model-OTP are shown in Figure~\ref{fig:modelotp_ssoall_windmodel} and are tabulated in Table~\ref{tab:modelotp_ssoall_windmodel}.

\begin{table*}
\center
\begin{tabular}{|l|l|lll|l|}
\cline{3-6}
\multicolumn{1}{c}{} & \multicolumn{1}{c|}{} & \multicolumn{3}{c|}{Bufton wind speed model (m/s)} & \multicolumn{1}{c|}{Wind direction} \\
\cline{2-5}
\multicolumn{1}{c|}{} & \multicolumn{1}{c|}{height (m)} & \multicolumn{1}{c|}{good} & \multicolumn{1}{c|}{typical} & \multicolumn{1}{c|}{bad} & \multicolumn{1}{c|}{(degrees, ref 0m)} \\
\hline
\multicolumn{1}{|c|}{GL} & \multicolumn{1}{c|}{37.5} & \multicolumn{1}{c}{2.1981} & \multicolumn{1}{c}{5.6605} & \multicolumn{1}{c|}{9.4200} & \multicolumn{1}{c|}{1.1674} \\
\multicolumn{1}{|c|}{} & \multicolumn{1}{c|}{250} & \multicolumn{1}{c}{2.2506} & \multicolumn{1}{c}{5.7810} & \multicolumn{1}{c|}{9.6142} & \multicolumn{1}{c|}{7.5347} \\
\hline
\multicolumn{1}{|c|}{FA} & \multicolumn{1}{c|}{1000} & \multicolumn{1}{c}{2.5495} & \multicolumn{1}{c}{6.3707} & \multicolumn{1}{c|}{10.4871} & \multicolumn{1}{c|}{26.9588} \\
\multicolumn{1}{|c|}{} & \multicolumn{1}{c|}{3000} & \multicolumn{1}{c}{5.1620} & \multicolumn{1}{c}{9.9300} & \multicolumn{1}{c|}{14.7605} & \multicolumn{1}{c|}{61.3717} \\
\multicolumn{1}{|c|}{} & \multicolumn{1}{c|}{6000} & \multicolumn{1}{c}{19.0935} & \multicolumn{1}{c}{23.4552} & \multicolumn{1}{c|}{27.9741} & \multicolumn{1}{c|}{85.7917} \\
\multicolumn{1}{|c|}{} & \multicolumn{1}{c|}{9000} & \multicolumn{1}{c}{32.0000} & \multicolumn{1}{c}{38.6276} & \multicolumn{1}{c|}{43.7936} & \multicolumn{1}{c|}{95.5085} \\
\multicolumn{1}{|c|}{} & \multicolumn{1}{c|}{13500} & \multicolumn{1}{c}{10.4619} & \multicolumn{1}{c}{26.4419} & \multicolumn{1}{c|}{41.6249} & \multicolumn{1}{c|}{100.3182} \\
\hline
\end{tabular}
\caption{\label{tab:modelotp_ssoall_windmodel} Tabulated values for the model wind speed and model wind direction for the GL and FA model-OTP layers for SSO (Run 1-8: May 2005 to June 2006) model.}
\end{table*}

\subsection{Summary Model-OTP}

The summary model-OTP (Kolmogorov) is the combination of the `good', `typical' and `bad' GL and FA model-OTPs resulting in a set of nine possible thin-layer turbulence profiles. Each turbulence profile of the model-OTP includes information about the turbulence layer strength, wind speed and wind direction.

A detailed summary table of the summary model-OTPs describing the observations at Siding Spring Observatory, Australia for the SSO (Run 1-8: May 2005 to June 2006) runs are listed in Table~\ref{tab:modelotp_ssoall_summary_integral} (turbulence integrals); Table~\ref{tab:modelotp_ssoall_summary_windspeeds} (turbulence layer wind speeds) and Table~\ref{tab:modelotp_ssoall_summary_winddirections} (turbulence layer wind direction). Note the model-OTP seeing, $\epsilon$, for the typical GL+FA of 1.24" is similar to historical median DIMM seeing of 1.25" reported by \cite{Wood1995}.

\begin{landscape}
\begin{table}[tbp]
\center
\begin{tabular}{|l|l|lll|lll|lll|}
\cline{3-11}
\multicolumn{1}{l}{} &  & \multicolumn{9}{c|}{Model Turbulence Profiles ($J$, Integral) - SSO (Run 1-8: May 2005 to June 2006)} \\
\cline{2-11}
\multicolumn{1}{l|}{} & GL & \multicolumn{3}{c|}{Good} & \multicolumn{3}{c|}{Typical} & \multicolumn{3}{c|}{Bad} \\
\cline{2-11}
\multicolumn{1}{l|}{} & FA & \multicolumn{1}{l|}{Good} & \multicolumn{1}{l|}{Typical} & Bad & \multicolumn{1}{l|}{Good} & \multicolumn{1}{l|}{Typical} & Bad & \multicolumn{1}{l|}{Good} & \multicolumn{1}{l|}{Typical} & Bad \\
\hline
Parameter & Units & \multicolumn{1}{l|}{1} & \multicolumn{1}{l|}{2} & 3 & \multicolumn{1}{l|}{4} & \multicolumn{1}{l|}{5} & 6 & \multicolumn{1}{l|}{7} & \multicolumn{1}{l|}{8} & 9 \\
\hline
37.5 & $10^{-13} m^{1/3}$ & 4.7129 & 4.7129 & 4.7129 & 7.3989 & 7.3989 & 7.3989 & 9.4363 & 9.4363 & 9.4363 \\
250 & $10^{-13} m^{1/3}$ & 0.2485 & 0.2485 & 0.2485 & 1.1574 & 1.1574 & 1.1574 & 3.8486 & 3.8486 & 3.8486 \\
1000 & $10^{-13} m^{1/3}$ & 0.2413 & 0.4772 & 0.6677 & 0.2413 & 0.4772 & 0.6677 & 0.2413 & 0.4772 & 0.6677 \\
3000 & $10^{-13} m^{1/3}$ & 0.0315 & 0.4368 & 1.1384 & 0.0315 & 0.4368 & 1.1384 & 0.0315 & 0.4368 & 1.1384 \\
6000 & $10^{-13} m^{1/3}$ & 0.0204 & 0.0924 & 0.3588 & 0.0204 & 0.0924 & 0.3588 & 0.0204 & 0.0924 & 0.3588 \\
9000 & $10^{-13} m^{1/3}$ & 0.0291 & 0.0863 & 0.2221 & 0.0291 & 0.0863 & 0.2221 & 0.0291 & 0.0863 & 0.2221 \\
13500 & $10^{-13} m^{1/3}$ & 0.0656 & 0.1187 & 0.3289 & 0.0656 & 0.1187 & 0.3289 & 0.0656 & 0.1187 & 0.3289 \\
\hline
$J_{GL}$ & $10^{-13} m^{1/3}$ & 4.9614 & 4.9614 & 4.9614 & 8.5562 & 8.5562 & 8.5562 & 13.2849 & 13.2849 & 13.2849 \\
$J_{FA}$ & $10^{-13} m^{1/3}$ & 0.3880 & 1.2115 & 2.7159 & 0.3880 & 1.2115 & 2.7159 & 0.3880 & 1.2115 & 2.7159 \\
$J$ & $10^{-13} m^{1/3}$ & 5.3493 & 6.1729 & 7.6773 & 8.9442 & 9.7678 & 11.2721 & 13.6728 & 14.4964 & 16.0008 \\
\hline
$F_{GL}$ & / 1.0 & 0.9275 & 0.8037 & 0.6462 & 0.9566 & 0.8760 & 0.7591 & 0.9716 & 0.9164 & 0.8303 \\
$F_{FA}$ & / 1.0 & 0.0725 & 0.1963 & 0.3538 & 0.0434 & 0.1240 & 0.2409 & 0.0284 & 0.0836 & 0.1697 \\
\hline
$\epsilon_{GL}$ & arcsecs & 0.8277 & 0.8277 & 0.8277 & 1.1478 & 1.1478 & 1.1478 & 1.4945 & 1.4945 & 1.4945 \\
$\epsilon_{FA}$  & arcsecs & 0.1794 & 0.3552 & 0.5766 & 0.1794 & 0.3552 & 0.5766 & 0.1794 & 0.3552 & 0.5766 \\
$\epsilon$ & arcsecs & 0.8659 & 0.9436 & 1.0755 & 1.1787 & 1.2427 & 1.3542 & 1.5206 & 1.5749 & 1.6710 \\
\hline
$\theta_0$ & arcsecs & 6.4233 & 3.7172 & 2.0123 & 6.3684 & 3.7043 & 2.0098 & 6.2255 & 3.6700 & 2.0030 \\
$\tau$ & ms & 11.7922 & 5.3516 & 2.3291 & 4.5855 & 3.5038 & 2.0310 & 2.2067 & 2.0112 & 1.5242 \\
\hline
Probability & / 1.0 & 0.0625 & 0.1250 & 0.0625 & 0.1250 & 0.2500 & 0.1250 & 0.0625 & 0.1250 & 0.0625 \\
\hline
\end{tabular}
\caption{\label{tab:modelotp_ssoall_summary_integral} Tabulated values for the final model-OTP for SSO (Run 1-8: May 2005 to June 2006), with layers specified as turbulence integral, $J$, in units $10^{-13} m^{1/3}$.}
\end{table}
\end{landscape}

\begin{landscape}
\begin{table}[tbp]
\center
\begin{tabular}{|l|l|lll|lll|lll|}
\cline{3-11}
\multicolumn{1}{l}{} &  & \multicolumn{9}{c|}{Model Turbulence Profiles (Layer wind speeds) - SSO (Run 1-8: May 2005 to June 2006)} \\
\cline{2-11}
\multicolumn{1}{l|}{} & GL & \multicolumn{3}{c|}{Good} & \multicolumn{3}{c|}{Typical} & \multicolumn{3}{c|}{Bad} \\
\cline{2-11}
\multicolumn{1}{l|}{} & FA & \multicolumn{1}{l|}{Good} & \multicolumn{1}{l|}{Typical} & Bad & \multicolumn{1}{l|}{Good} & \multicolumn{1}{l|}{Typical} & Bad & \multicolumn{1}{l|}{Good} & \multicolumn{1}{l|}{Typical} & Bad \\
\hline
Parameter & Units & \multicolumn{1}{l|}{1} & \multicolumn{1}{l|}{2} & 3 & \multicolumn{1}{l|}{4} & \multicolumn{1}{l|}{5} & 6 & \multicolumn{1}{l|}{7} & \multicolumn{1}{l|}{8} & 9 \\
\hline
37.5 & m/s & 2.1981 & 2.1981 & 2.1981 & 5.6605 & 5.6605 & 5.6605 & 9.4200 & 9.4200 & 9.4200 \\
250 & m/s & 2.2506 & 2.2506 & 2.2506 & 5.7810 & 5.7810 & 5.7810 & 9.6142 & 9.6142 & 9.6142 \\
1000 & m/s & 2.5495 & 6.3707 & 10.4871 & 2.5495 & 6.3707 & 10.4871 & 2.5495 & 6.3707 & 10.4871 \\
3000 & m/s & 5.1620 & 9.9300 & 14.7605 & 5.1620 & 9.9300 & 14.7605 & 5.1620 & 9.9300 & 14.7605 \\
6000 & m/s & 19.0935 & 23.4552 & 27.9741 & 19.0935 & 23.4552 & 27.9741 & 19.0935 & 23.4552 & 27.9741 \\
9000 & m/s & 32.0000 & 38.6276 & 43.7936 & 32.0000 & 38.6276 & 43.7936 & 32.0000 & 38.6276 & 43.7936 \\
13500 & m/s & 10.4619 & 26.4419 & 41.6250 & 10.4619 & 26.4419 & 41.6250 & 10.4619 & 26.4419 & 41.6250 \\
\hline
$J_{GL}$ & $10^{-13} m^{1/3}$ & 4.9614 & 4.9614 & 4.9614 & 8.5562 & 8.5562 & 8.5562 & 13.2849 & 13.2849 & 13.2849 \\
$J_{FA}$ & $10^{-13} m^{1/3}$ & 0.3880 & 1.2115 & 2.7159 & 0.3880 & 1.2115 & 2.7159 & 0.3880 & 1.2115 & 2.7159 \\
$J$ & $10^{-13} m^{1/3}$ & 5.3493 & 6.1729 & 7.6773 & 8.9442 & 9.7678 & 11.2721 & 13.6728 & 14.4964 & 16.0008 \\
\hline
$F_{GL}$ & / 1.0 & 0.9275 & 0.8037 & 0.6462 & 0.9566 & 0.8760 & 0.7591 & 0.9716 & 0.9164 & 0.8303 \\
$F_{FA}$ & / 1.0 & 0.0725 & 0.1963 & 0.3538 & 0.0434 & 0.1240 & 0.2409 & 0.0284 & 0.0836 & 0.1697 \\
\hline
$\epsilon_{GL}$ & arcsecs & 0.8277 & 0.8277 & 0.8277 & 1.1478 & 1.1478 & 1.1478 & 1.4945 & 1.4945 & 1.4945 \\
$\epsilon_{FA}$  & arcsecs & 0.1794 & 0.3552 & 0.5766 & 0.1794 & 0.3552 & 0.5766 & 0.1794 & 0.3552 & 0.5766 \\
$\epsilon$ & arcsecs & 0.8659 & 0.9436 & 1.0755 & 1.1787 & 1.2427 & 1.3542 & 1.5206 & 1.5749 & 1.6710 \\
\hline
$\theta_0$ & arcsecs & 6.4233 & 3.7172 & 2.0123 & 6.3684 & 3.7043 & 2.0098 & 6.2255 & 3.6700 & 2.0030 \\
$\tau$ & ms & 11.7922 & 5.3516 & 2.3291 & 4.5855 & 3.5038 & 2.0310 & 2.2067 & 2.0112 & 1.5242 \\
\hline
Probability & / 1.0 & 0.0625 & 0.1250 & 0.0625 & 0.1250 & 0.2500 & 0.1250 & 0.0625 & 0.1250 & 0.0625 \\
\hline
\end{tabular}
\caption{\label{tab:modelotp_ssoall_summary_windspeeds} Tabulated values for the final model-OTP for the SSO (Run 1-8: May 2005 to June 2006), with layers specified as wind speeds.}
\end{table}
\end{landscape}

\begin{landscape}
\begin{table}[tbp]
\center
\begin{tabular}{|l|l|lll|lll|lll|}
\cline{3-11}
\multicolumn{1}{l}{} &  & \multicolumn{9}{c|}{Model Turbulence Profiles (Layer wind directions) - SSO (Run 1-8: May 2005 to June 2006)} \\
\cline{2-11}
\multicolumn{1}{l|}{} & GL & \multicolumn{3}{c|}{Good} & \multicolumn{3}{c|}{Typical} & \multicolumn{3}{c|}{Bad} \\
\cline{2-11}
\multicolumn{1}{l|}{} & FA & \multicolumn{1}{l|}{Good} & \multicolumn{1}{l|}{Typical} & Bad & \multicolumn{1}{l|}{Good} & \multicolumn{1}{l|}{Typical} & Bad & \multicolumn{1}{l|}{Good} & \multicolumn{1}{l|}{Typical} & Bad \\
\hline
Parameter & Units & \multicolumn{1}{l|}{1} & \multicolumn{1}{l|}{2} & 3 & \multicolumn{1}{l|}{4} & \multicolumn{1}{l|}{5} & 6 & \multicolumn{1}{l|}{7} & \multicolumn{1}{l|}{8} & 9 \\
\hline
37.5 & degrees & 1.1674 & 1.1674 & 1.1674 & 1.1674 & 1.1674 & 1.1674 & 1.1674 & 1.1674 & 1.1674 \\
250 & degrees & 7.5347 & 7.5347 & 7.5347 & 7.5347 & 7.5347 & 7.5347 & 7.5347 & 7.5347 & 7.5347 \\
1000 & degrees & 26.9588 & 26.9588 & 26.9588 & 26.9588 & 26.9588 & 26.9588 & 26.9588 & 26.9588 & 26.9588 \\
3000 & degrees & 61.3717 & 61.3717 & 61.3717 & 61.3717 & 61.3717 & 61.3717 & 61.3717 & 61.3717 & 61.3717 \\
6000 & degrees & 85.7917 & 85.7917 & 85.7917 & 85.7917 & 85.7917 & 85.7917 & 85.7917 & 85.7917 & 85.7917 \\
9000 & degrees & 95.5085 & 95.5085 & 95.5085 & 95.5085 & 95.5085 & 95.5085 & 95.5085 & 95.5085 & 95.5085 \\
13500 & degrees & 100.3182 & 100.3182 & 100.3182 & 100.3182 & 100.3182 & 100.3182 & 100.3182 & 100.3182 & 100.3182 \\
\hline
$J_{GL}$ & $10^{-13} m^{1/3}$ & 4.9614 & 4.9614 & 4.9614 & 8.5562 & 8.5562 & 8.5562 & 13.2849 & 13.2849 & 13.2849 \\
$J_{FA}$ & $10^{-13} m^{1/3}$ & 0.3880 & 1.2115 & 2.7159 & 0.3880 & 1.2115 & 2.7159 & 0.3880 & 1.2115 & 2.7159 \\
$J$ & $10^{-13} m^{1/3}$ & 5.3493 & 6.1729 & 7.6773 & 8.9442 & 9.7678 & 11.2721 & 13.6728 & 14.4964 & 16.0008 \\
\hline
$F_{GL}$ & / 1.0 & 0.9275 & 0.8037 & 0.6462 & 0.9566 & 0.8760 & 0.7591 & 0.9716 & 0.9164 & 0.8303 \\
$F_{FA}$ & / 1.0 & 0.0725 & 0.1963 & 0.3538 & 0.0434 & 0.1240 & 0.2409 & 0.0284 & 0.0836 & 0.1697 \\
\hline
$\epsilon_{GL}$ & arcsecs & 0.8277 & 0.8277 & 0.8277 & 1.1478 & 1.1478 & 1.1478 & 1.4945 & 1.4945 & 1.4945 \\
$\epsilon_{FA}$  & arcsecs & 0.1794 & 0.3552 & 0.5766 & 0.1794 & 0.3552 & 0.5766 & 0.1794 & 0.3552 & 0.5766 \\
$\epsilon$ & arcsecs & 0.8659 & 0.9436 & 1.0755 & 1.1787 & 1.2427 & 1.3542 & 1.5206 & 1.5749 & 1.6710 \\
\hline
$\theta_0$ & arcsecs & 6.4233 & 3.7172 & 2.0123 & 6.3684 & 3.7043 & 2.0098 & 6.2255 & 3.6700 & 2.0030 \\
$\tau$ & ms & 11.7922 & 5.3516 & 2.3291 & 4.5855 & 3.5038 & 2.0310 & 2.2067 & 2.0112 & 1.5242 \\
\hline
Probability & / 1.0 & 0.0625 & 0.1250 & 0.0625 & 0.1250 & 0.2500 & 0.1250 & 0.0625 & 0.1250 & 0.0625 \\
\hline
\end{tabular}
\caption{\label{tab:modelotp_ssoall_summary_winddirections} Tabulated values for the final model-OTP for the SSO (Run 1-8: May 2005 to June 2006), with layers specified as wind directions.}
\end{table}
\end{landscape}

%%%%%%%%%%%%%%%%%%%%

\section{Conclusions}

This paper has reported on our turbulence profiling observational results performed at SSO.  A summary of results with example data has been presented for measurement spanning years 2005 to 2006 using the  $7\times7$ (Runs 1-6) and $17\times17$ (Runs 7-8) SLODAR instruments. The observational results has facilitated the site-characterisation of the optical turbulence profile with the implementation of model-OTP (Kolmogorov) for SSO. The model-OTP describes the turbulence layer strength (Table~\ref{tab:modelotp_ssoall_summary_integral}), layer wind speed  (Table~\ref{tab:modelotp_ssoall_summary_windspeeds}) and layer wind direction (Table~\ref{tab:modelotp_ssoall_summary_winddirections}). The model-OTP is useful for the prediction of adaptive optics performance at SSO (forthcoming paper) using simulation codes. Prior to the commencement of our SLODAR campaign the seeing statistics were relatively well understood with DIMM seeing measurements \citep{Wood1995}, whereas the vertical structure of the turbulence profile structure was relatively unknown, with only a few SCIDAR profiles reported by \cite{Klueckers1998}. The following conclusions can be stated about the atmospheric turbulence above SSO based on our data:

\begin{itemize}
  \item A measured median atmospheric seeing of around 1.2" (Kolmogorov model with mirror/dome seeing removed). The seeing is discussed in Section~\ref{sec:turbulencedistribution}. 

  \item Ground-layer turbulence dominates, with $\sim$80\% turbulence below 500~m ($h_{500}$). The structure of the turbulence is discussed in Section~\ref{sec:turbulencedistribution}. The presence of a strong ground-layer was also discovered during site measurements at Mount John University Observatory~\citep{Mohr2010}. The dominate ground-layer has promising implications for  GLAO. A forthcoming paper will investigate the performance of GLAO based on the model-OTP for SSO.  

  \item Non-Kolmogorov turbulence is observed especially for ground-layer, with $\beta_{avg}\sim10/3$. The power-law slope of the spatial phase fluctuations is discussed in Section~\ref{sec:turbulencedistribution}. The non-Kolomogorov spectrum  results in a different scaling of seeing with wavelength to that conventionally assumed.

  \item Turbulence profiles shows a number of dynamical characteristics; (i) most intense  near the ground (below 100~m); (ii)  fluctuate in intensity, appearing in `bursts' with timescales of several minutes. These characteristics are shown in Figure~\ref{sso_indv_profile_run7} as a sequence of turbulence profiles.

  \item Mirror/dome seeing can be a significant fraction of the ground-layer turbulence. This is evident as a zero-height static contribution in the turblent layer wind speed measurements (refer to Section~\ref{sec:layerwindprofiles} and Figure~\ref{fig:sso_wind_run7}).

  \item The free-atmosphere turbulence is comparable to 'good' seeing sites. From the model-OTP shown in Table~\ref{tab:modelotp_ssoall_summary_integral}, the free-atmosphere seeing for SSO is 0.18 (25\%), 0.36 (50\%) and 0.58 (75\%) arcsecs. The Cerro Pachon (seeing$\sim$0.75)  model (CP Model, \cite{TokovininTravouillon2006}) for the free-atmosphere reports 0.29 (25\%), 0.40 (50\%) and 0.55 (75\%) arcsecs.
\end{itemize}

%% It is preferable to embed your figures in the text as in the following example
%\begin{figure}[]
%\begin{center}
%%\includegraphics[scale=1, angle=0]{figure.eps}
%\caption{An example figure caption.}\label{figexample}
%\end{center}
%\end{figure}
%
%
%
%%%Format tables as in the following example
%\begin{table}[h]
%\begin{center}
%\caption{Example Table}\label{tableexample}
%\begin{tabular}{lcc}
%\hline Column 1 & Column 2 & Column 3 \\
%\hline Table Content$^a$ \\
%\hline
%\end{tabular}
%\medskip\\
%$^a$Table footnotes go here.\\
%\end{center}
%\end{table}

\section*{Acknowledgments} %If needed

This work makes use of observational data and analysis provided by the author's PhD thesis research (submitted 2009) conducted at the  Research School of Astronomy and Astrophysics (RSAA) of the  Australian National University (ANU). We thank Peter Conroy (ANU) for the design and manufacture of the instrumentation. The authors are grateful to Jon Lawrence of the Australian Astronomical Observatory (AAO) for his suggestions regarding the original version of the manuscript and to the referees for their comments.

%\begin{thebibliography}{}
%% References are listed as in the following example, for more examples, please
%% see the PASA Style Guide
%\bibitem[Smith, Jones, \& Brown(Year)Smith et al.]{example}Smith, A.~B., Jones,
%C.~D., Brown, E.~F. Year, Journal, Volume, Page
%\end{thebibliography}

\bibliographystyle{astron} % apj % astron
\bibliography{references}

\begin{thebibliography}{}

\bibitem[\protect\astroncite{{Abahamid} et~al.}{2004}]{Abahamid2004}
{Abahamid}, A., {Jabiri}, A., {Vernin}, J., {Benkhaldoun}, Z., {Azouit}, M.,
  and {Agabi}, A.: 2004,
\newblock {\em \aap} {\bf 416}, 1193

\bibitem[\protect\astroncite{{Andersen} et~al.}{2006}]{Andersen2006}
{Andersen}, D.~R., {Stoesz}, J., {Morris}, S., {Lloyd-Hart}, M., {Crampton},
  D., {Butterley}, T., {Ellerbroek}, B., {Jolissaint}, L., {Milton}, N.~M.,
  {Myers}, R., {Szeto}, K., {Tokovinin}, A., {V{\'e}ran}, J.-P., and {Wilson},
  R.: 2006,
\newblock {\em \pasp} {\bf 118}, 1574

\bibitem[\protect\astroncite{{Avila} et~al.}{2003}]{Avila2003}
{Avila}, R., {Iba{\~n}ez}, F., {Vernin}, J., {Masciadri}, E., {S{\'a}nchez},
  L.~J., {Azouit}, M., {Agabi}, A., {Cuevas}, S., and {Garfias}, F.: 2003,
\newblock in I. {Cruz-Gonzalez}, R. {Avila}, and M. {Tapia} (eds.), {\em
  Revista Mexicana de Astronomia y Astrofisica Conference Series}, Vol.~19 of
  {\em Revista Mexicana de Astronomia y Astrofisica, vol. 27}, pp 11--22

\bibitem[\protect\astroncite{{Azouit} and {Vernin}}{2005}]{Azouit2005}
{Azouit}, M. and {Vernin}, J.: 2005,
\newblock {\em \pasp} {\bf 117}, 536

\bibitem[\protect\astroncite{{Butterley} et~al.}{2006}]{Butterley2006}
{Butterley}, T., {Wilson}, R.~W., and {Sarazin}, M.: 2006,
\newblock {\em \mnras} {\bf 369}, 835

\bibitem[\protect\astroncite{{Ellerbroek} and {Rigaut}}{2000}]{Ellerbroek2000}
{Ellerbroek}, B.~L. and {Rigaut}, F.~J.: 2000,
\newblock in P.~L. {Wizinowich} (ed.), {\em Society of Photo-Optical
  Instrumentation Engineers (SPIE) Conference Series}, Vol. 4007 of {\em
  Presented at the Society of Photo-Optical Instrumentation Engineers (SPIE)
  Conference}, pp 1088--1099

\bibitem[\protect\astroncite{{Fried}}{1966}]{Fried1966}
{Fried}, D.~L.: 1966,
\newblock {\em Journal of the Optical Society of America (1917-1983)} {\bf 56},
  1372

\bibitem[\protect\astroncite{{Fried}}{1982}]{Fried1982}
{Fried}, D.~L.: 1982,
\newblock {\em Journal of the Optical Society of America (1917-1983)} {\bf 72},
  52

\bibitem[\protect\astroncite{{Fuchs} et~al.}{1994}]{Fuchs1994}
{Fuchs}, A., {Tallon}, M., and {Vernin}, J.: 1994,
\newblock in W.~A. {Flood} and W.~B. {Miller} (eds.), {\em Proc. SPIE Vol.
  2222, p. 682-692, Atmospheric Propagation and Remote Sensing III, Walter A.
  Flood; Walter B. Miller; Eds.}, Vol. 2222 of {\em Presented at the Society of
  Photo-Optical Instrumentation Engineers (SPIE) Conference}, pp 682--692

\bibitem[\protect\astroncite{{Goodwin}}{2009}]{Goodwin2009}
{Goodwin}, M.: 2009,
\newblock {\em {Turbulence profiling at Siding Spring and Las Campanas
  Observatories}},
\newblock ANU PhD Thesis

\bibitem[\protect\astroncite{{Goodwin} et~al.}{2007}]{Goodwin2007}
{Goodwin}, M., {Jenkins}, C., and {Lambert}, A.: 2007,
\newblock {\em Optics Express} {\bf 151}, 14844

\bibitem[\protect\astroncite{{Greenwood}}{1977}]{Greenwood1977}
{Greenwood}, D.~P.: 1977,
\newblock {\em Journal of the Optical Society of America (1917-1983)} {\bf 67},
  390

\bibitem[\protect\astroncite{{Hardy}}{1998}]{Hardy1998}
{Hardy}, J.~W.: 1998,
\newblock {\em {Adaptive Optics for Astronomical Telescopes}},
\newblock Adaptive Optics for Astronomical Telescopes, by John W Hardy,
  pp.~448.~Foreword by John W Hardy.~Oxford University Press, Jul
  1998.~ISBN-10: 0195090195.~ISBN-13: 9780195090192

\bibitem[\protect\astroncite{{Hubin} et~al.}{2006}]{Hubin2006}
{Hubin}, N., {Ellerbroek}, B.~L., {Arsenault}, R., {Clare}, R.~M., {Dekany},
  R., {Gilles}, L., {Kasper}, M., {Herriot}, G., {Le Louarn}, M., {Marchetti},
  E., {Oberti}, S., {Stoesz}, J., {Veran}, J.~P., and {V{\'e}rinaud}, C.: 2006,
\newblock in P. {Whitelock}, M. {Dennefeld}, and B. {Leibundgut} (eds.), {\em
  The Scientific Requirements for Extremely Large Telescopes}, Vol. 232 of {\em
  IAU Symposium}, pp 60--85

\bibitem[\protect\astroncite{{JAI Inc}}{2009}]{PulnixRef}
{JAI Inc}: 2009,
\newblock {\em {Pulnix}},
\newblock \url{http://www.jai.com/}

\bibitem[\protect\astroncite{{Klueckers} et~al.}{1998}]{Klueckers1998}
{Klueckers}, V.~A., {Wooder}, N.~J., {Nicholls}, T.~W., {Adcock}, M.~J.,
  {Munro}, I., and {Dainty}, J.~C.: 1998,
\newblock {\em \aaps} {\bf 130}, 141

\bibitem[\protect\astroncite{MathWorks}{2005}]{Matlabref}
MathWorks: 2005,
\newblock {\em {MATLAB v7.0.4.365 (R14) Service Pack 2}},
\newblock \url{http://www.mathworks.com/products/matlab/}

\bibitem[\protect\astroncite{{Mohr} et~al.}{2010}]{Mohr2010}
{Mohr}, J.~L., {Johnston}, R.~A., and {Cottrell}, P.~L.: 2010,
\newblock {\em \pasa} {\bf 27}, 347

\bibitem[\protect\astroncite{{Pant} et~al.}{1999}]{Pant1999}
{Pant}, P., {Stalin}, C.~S., and {Sagar}, R.: 1999,
\newblock {\em \aaps} {\bf 136}, 19

\bibitem[\protect\astroncite{PixeLINK}{2009}]{PixelinkRef}
PixeLINK: 2009,
\newblock {\em {Pixelink}},
\newblock \url{http://www.pixelink.com/}

\bibitem[\protect\astroncite{Sarazin et~al.}{2005}]{Sarazin2005}
Sarazin, M., Butterley, T., Tokovinin, A., Travouillon, T., and Wilson, R.:
  2005,
\newblock {\em The Tololo SLODAR Campaign, Final Report},
\newblock \url{https://www.eso.org/genfac/pubs/astclim/paranal/asm/slodar/The
  Tololo SLODAR Campaign.htm}

\bibitem[\protect\astroncite{{Stribling} et~al.}{1995}]{Stribling1995}
{Stribling}, B.~E., {Welsh}, B.~M., and {Roggemann}, M.~C.: 1995,
\newblock in J.~C. {Dainty} (ed.), {\em Proc. SPIE Vol. 2471, p. 181-196,
  Atmospheric Propagation and Remote Sensing IV, J. Christopher Dainty; Ed.},
  Vol. 2471 of {\em Presented at the Society of Photo-Optical Instrumentation
  Engineers (SPIE) Conference}, pp 181--196

\bibitem[\protect\astroncite{{Tokovinin} et~al.}{2003}]{Tokovinin2003b}
{Tokovinin}, A., {Baumont}, S., and {Vasquez}, J.: 2003,
\newblock {\em \mnras} {\bf 340}, 52

\bibitem[\protect\astroncite{{Tokovinin} and
  {Travouillon}}{2006}]{TokovininTravouillon2006}
{Tokovinin}, A. and {Travouillon}, T.: 2006,
\newblock {\em \mnras} {\bf 365}, 1235

\bibitem[\protect\astroncite{{Travouillon}}{2006}]{Travouillon2006}
{Travouillon}, T.: 2006,
\newblock in {\em Ground-based and Airborne Telescopes. Edited by Stepp, Larry
  M.. Proceedings of the SPIE, Volume 6267, pp. 626720 (2006).}, Vol. 6267 of
  {\em Presented at the Society of Photo-Optical Instrumentation Engineers
  (SPIE) Conference}

\bibitem[\protect\astroncite{{Vernin} et~al.}{1998}]{Vernin1998}
{Vernin}, J., {Agabi}, A., {Avila}, R., {Azouit}, M., {Conan}, R., {Martin},
  F., {Masciadri}, E., {Sanchez}, L., and {Ziad}, A.: 1998,
\newblock {\em {1998 Gemini site testing campaign. Cerro Pachon and Cerro
  Tololo}},
\newblock Gemini Report RPT-AO-G0094

\bibitem[\protect\astroncite{{Vernin} and {Roddier}}{1973}]{Vernin1973}
{Vernin}, J. and {Roddier}, F.: 1973,
\newblock {\em Journal of the Optical Society of America (1917-1983)} {\bf 63},
  270

\bibitem[\protect\astroncite{{Wilson}}{2002}]{Wilson2002}
{Wilson}, R.~W.: 2002,
\newblock {\em \mnras} {\bf 337}, 103

\bibitem[\protect\astroncite{{Wood} et~al.}{1995}]{Wood1995}
{Wood}, P.~R., {Rodgers}, A.~W., and {Russell}, K.~S.: 1995,
\newblock {\em Publications of the Astronomical Society of Australia} {\bf 12},
  97

\end{thebibliography}

%\end{multicols}

\end{document}